\documentclass[]{wiley-article}
\usepackage{natbib}
\usepackage{siunitx}
\usepackage{float}
\usepackage{longtable}
\usepackage{numprint}
\usepackage{eurosym}
\usepackage[hidelinks]{hyperref}
\usepackage{tikz} 
\usepackage[boxed]{algorithm2e}
\usepackage[font=small,labelfont=bf]{subcaption}
\usepackage{caption}
\definecolor{myorange}{RGB}{255, 162, 8}
\definecolor{myyellow}{RGB}{250, 231, 0}
\definecolor{mypink}{RGB}{255, 85, 247}
\definecolor{myred}{RGB}{255, 0, 0}
\definecolor{myviolet}{RGB}{148, 0, 211}

\papertype{Original Article}

\title{Automated Detection of Missing Links\\in Bicycle Networks}

\author[1]{Anastassia Vybornova}
\author[1]{Tiago Cunha}
\author[2]{\\Astrid Gühnemann}
\author[1,3,4]{Michael Szell}

\affil[1]{NEtwoRks, Data, and Society (NERDS), Computer Science Department, IT University of Copenhagen, 2300 Copenhagen, Denmark}
\affil[2]{Institute for Transport Studies,  University of Natural Resources and Life Sciences, 1180 Vienna, Austria}
\affil[3]{ISI Foundation, 10126 Turin, Italy}
\affil[4]{Complexity Science Hub Vienna, 1080 Vienna, Austria}

\corraddress{Michael Szell, Computer Science Department, IT University of Copenhagen, 2300 Copenhagen, Denmark}
\corremail{misz@itu.dk}

\fundinginfo{Danish Ministry of Transport, Road Directorate}

\runningauthor{VYBORNOVA et al.}

\begin{document}

\begin{frontmatter}
\maketitle

\begin{abstract}
Cycling is an effective solution for making urban transport more sustainable. However, bicycle networks are typically developed in a slow, piecewise process that leaves open a large number of gaps, even in well developed cycling cities like Copenhagen. Here, we develop the IPDC procedure (Identify, Prioritize, Decluster, Classify) for finding the most important missing links in urban bicycle networks, using data from OpenStreetMap. In this procedure we first identify all possible gaps following a multiplex network approach, prioritize them according to a flow-based metric, decluster emerging gap clusters, and manually classify the types of gaps. We apply the IPDC procedure to Copenhagen and report the 105 top priority gaps. For evaluation, we compare these gaps with the city's most recent Cycle Path Prioritization Plan and find considerable overlaps. Our results show how network analysis with minimal data requirements can serve as a cost-efficient support tool for bicycle network planning. By taking into account the whole city network for consolidating urban bicycle infrastructure, our data-driven framework can complement localized, manual planning processes for more effective, city-wide decision-making.

\end{abstract}
\end{frontmatter}

\section*{Introduction}

\begin{figure}[t]
    \centering
    \includegraphics[width=0.8\textwidth]{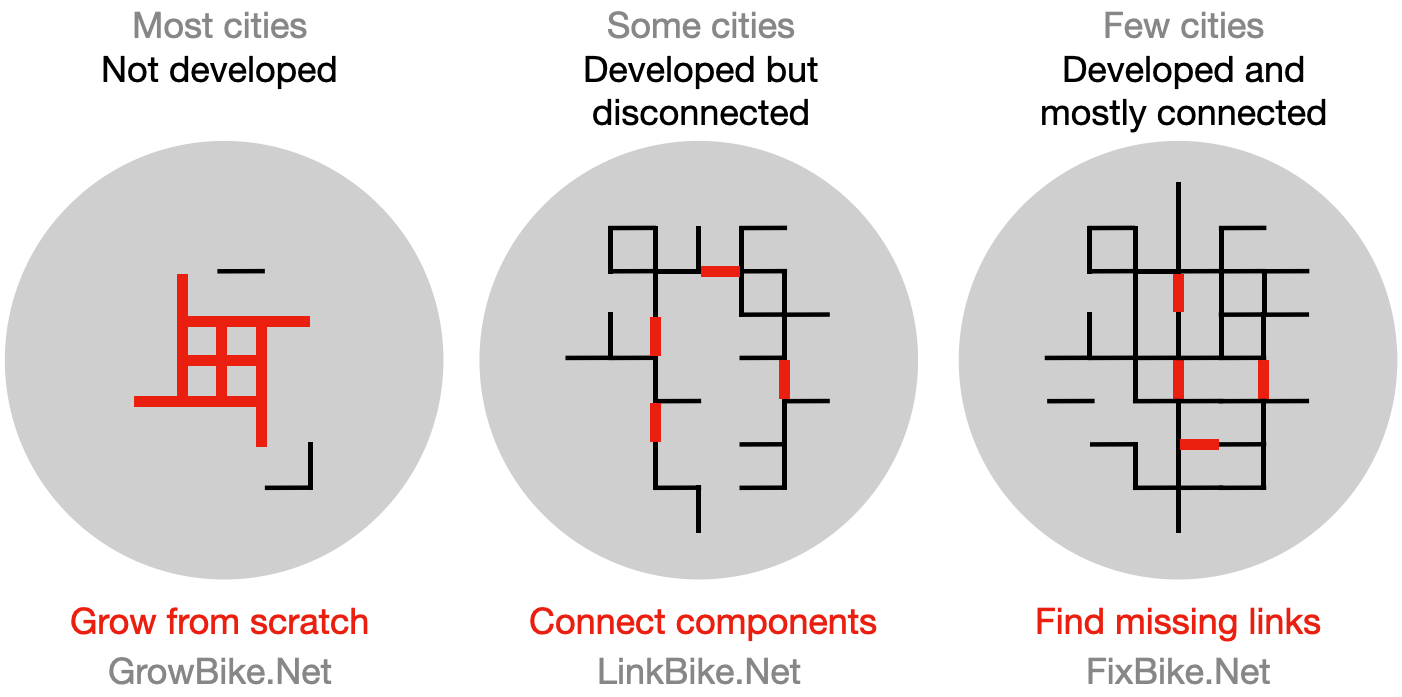}  
    \caption{\textbf{Depending on a city's existing bicycle network (black), different development approaches (red) can fit.} Left: The approach of growing from scratch by \cite{szell_growing_2021} is best applicable for underdeveloped cities, such as Los Angeles. See also: \url{https://growbike.net}. Center: The approach of \cite{nateraorozco2020dso} to connect disconnected components can well fit cities that have developed but disconnected components, such as Budapest. See also: \url{http://linkbike.net}. Right: Here we develop a process for finding missing links also within a connected component. This method complements the other two approaches; also it fits well cities with a developed and connected network such as Copenhagen. See also: \url{https://fixbike.net}.\label{fig:approaches}}
\end{figure}

With transport being one of the most problematic sectors in terms of emission reductions \citep{lamb2021rtd}, urban transportation systems play a decisive role in tackling the climate crisis. There is enormous potential to be harnessed by ``greening'' the transportation sector through a modal shift towards active and more sustainable mobility modes such as cycling and walking, both in terms of climate change mitigation and socioeconomic benefits \citep{gossling2019sca, high-level_advisory_group_on_sustainable_transport_mobilizing_2016}.

In practice, however, bicycle infrastructure development struggles with a particularly pervasive political inertia due to the complex interdependencies of car-centrism \citep{mattioli2020political,feddes2020hard}. Very few cities have so far managed to build up relatively safe and cohesive bicycle networks \citep{crow2016dmb}, and even the most renowned cycling cities in the world still have a long way to go to achieve a sustainable urban transport system, and an optimal cycling network. For instance, this is the case for Copenhagen, where despite over a century of political struggles and coordinated efforts to develop a functioning grid of protected on-street bicycle networks \citep{carstensen2015sdc}, its network of protected bicycle infrastructure is split into 300 disconnected components \citep{nateraorozco2020dso} and its accessibility displays considerable local variations \citep{rahbek_viero_connectivity_2020}. For the assessment of an urban bicycle network, it is therefore crucial to ask: ``Where are the missing links?'', ``How to fix them?'', and ``How much will this cost and benefit the city?'' These are the questions we aim to answer in this paper. Our approach is based on \cite{vybornova2021icg} to develop a generally applicable, computational procedure for finding missing links in developed bicycle networks, and testing it on the case of Copenhagen.

From a research perspective, a structured, data-driven approach to bicycle network planning, along with a strong theoretical and computational underpinning, is largely missing. Setting up such an approach is seen by many as necessary precondition for an evidence-based modal shift towards increased bicycle use and reduced car use \citep{koglin_marginalisation_2014, crow2016dmb, buehler_bikeway_2016, resch2019hds, priya_uteng_addressing_2019}. From this viewpoint, the academic literature on network analysis approaches to bicycle network planning can be divided into three broad categories, depending on the structure and reproducibility of the underlying approach. 

The first, largest category contains transport planning studies with a \emph{place-specific} focus. These case studies focus on improving the bicycle infrastructure of one particular city, for example, Seattle \citep{lowry2017qbn}, Toronto \citep{mitra_mode_2017}, Lisbon \citep{abad_quantifying_2018} or London \citep{palominos2021ica}. Characteristic for these studies is the specific application to one city and its idiosyncrasies, using a variety of data sets, such as orography, traffic flows, trip tables or citizen surveys on mobility preferences. The second, more recent approach, is based on the physics-inspired Science of Cities \citep{batty2013nsc} and aims to identify the generalized laws and mechanisms that govern urban development and are \emph{independent of place}. This approach typically focuses on the most important ``first-order'' effects following the paradigm of network science, sacrificing specificity for generality, and therefore deliberately using maximally simplified data sets. Given that this second approach aims for general results, it must be tested for multiple cities. Examples include a multiplex network study of multimodality \citep{nateraorozco2020emf}, methods to prioritize pop-up active transport infrastructure \citep{lovelace2020mpp}, linking disconnected components \citep{nateraorozco2020dso}, or growing bicycle networks from scratch \citep{szell_growing_2021}. Finally, the third category contains studies that develop \textit{generalizable} approaches based on the use case of one specific city \citep{larsen_build_2013, zhang_prioritizing_2014, boisjoly2020bnp, olmos2020dcf, reggiani_understanding_2021}. There is an inherent feasibility trade-off between developing a refined model by working with one high resolution data set versus developing a generalizable model by working with several lower resolution data sets. Studies from the third category therefore often imply a call to the cycling research community to collaborate on method consolidation by further testing their respective approach for other cities. 

Our approach developed here corresponds to the third category: we first develop a generalizable method for the detection of gaps in bicycle networks, and then carry out a detailed evaluation procedure for the use case of Copenhagen in order to demonstrate the applicability of our method. Our procedure should be applicable to other cities without major adjustments. Further, this new procedure could also be applied to less developed networks, for example to complement previous approaches (Fig.~\ref{fig:approaches}) or to find missing links in sub-networks below the scale of the city.

Lastly, there are also numerous approaches to bicycle network planning that focus on (actual or estimated) travel demand, often rooted in transport modeling \citep{dill_understanding_2008, lovelace2017pct, cooper_predictive_2018, skov-petersen_how_2018, van_eldijk_missing_2020}. The approaches divided in three categories above, in contrast, are all rooted in network analysis and focus on improvements of existing infrastructure, which our study is also in line with. 

Our method complements \cite{nateraorozco2020dso} and \cite{szell_growing_2021}, see Fig.~\ref{fig:approaches}: Instead of providing optimized improvements to cities with minimal existing networks \citep{szell_growing_2021}, or to cities with developed but still quite disconnected networks \citep{nateraorozco2020dso}, here we focus on repairing networks. This new approach is particularly suited for well developed networks in which the largest connected components already cover the majority of nodes. These networks do not benefit from an approach that starts from scratch. They can benefit from connecting existing components \citep{nateraorozco2020dso}, but since they cover already most of the city, this benefit becomes exhausted quickly once the few biggest components have been linked up. However, there can still be many missing links left \emph{within} their connected components, for which we set up an automated fixing procedure here, see Fig.~\ref{fig:workflow}. 

\section*{The IPDC procedure}

\begin{figure}[t]
    \centering
    \includegraphics[width=\textwidth]{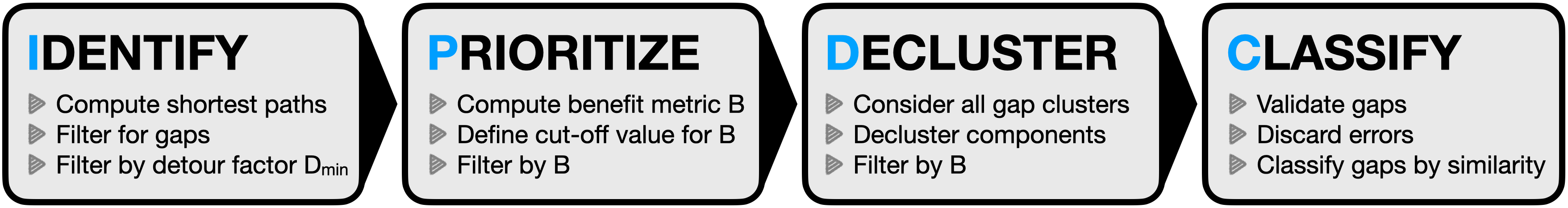}
    \caption{\textbf{Steps in the IPDC procedure.} Gaps are first identified via shortest paths, discarding parallel paths using a minimum detour factor $D_{\mathrm{min}}$. Gaps are then prioritized via a gap closure benefit metric, in the simplest case based on betweenness centrality. Resulting gaps can overlap (cluster) and need to be declustered. Finally, gaps are compared with existing infrastucture, validated or discarded, and classified.}
    \label{fig:workflow}
\end{figure}

A cyclist on their way through an urban bicycle network will often find themselves suddenly having to share the road with cars for a while, or having to cross unprotected intersections with a high traffic load, even in a well-connected, developed bicycle network like in Copenhagen. Here we formalize this intuitive concept of a ``missing link'' in the bicycle network and develop an automated procedure to find the most important ones. We call our procedure \emph{IPDC} after its four main steps: Identify, Prioritize, Decluster, Classify, which we present in this section. The IPDC procedure is illustrated in the workflow diagram in Fig.~\ref{fig:workflow}, and described in detail in the sections below. See Section \emph{Scope and limitations} for details on the applicability and limitations of this approach. We start by outlining the network data structure and our formal definition of ``gap'' used for the first step of the IPDC procedure, gap identification.

\subsection*{Gap identification}
\begin{figure}[t]
    \centering
    \includegraphics[width=0.85\textwidth]{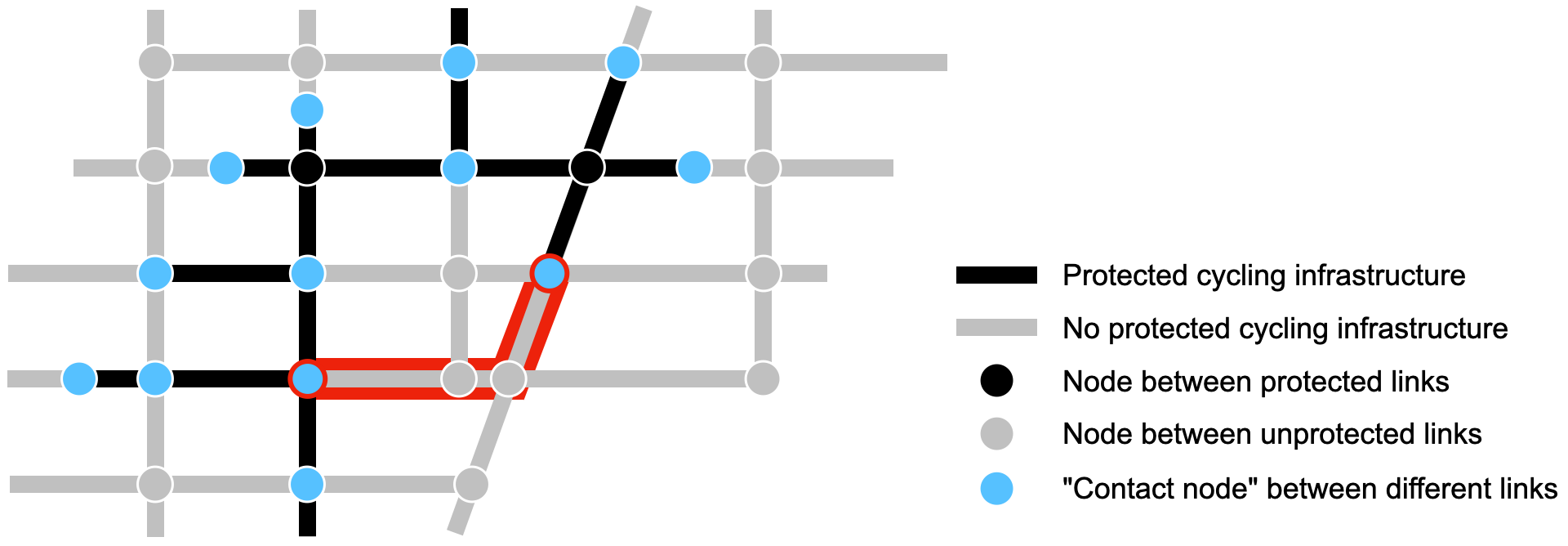}
\caption{\textbf{Illustration of node and link types, and of our definition of gap.} We define a gap as a shortest path between two contact nodes that consists only of unprotected links. An example of a gap between the two highlighted contact nodes is illustrated in red. (Here we only illustrate one out of many possible gaps between all pairs of contact nodes.)}
    \label{fig:gapdefinition}
\end{figure}

As a starting point, the IPDC procedure takes an urban network of streets and protected bicycle tracks, as provided by OpenStreetMap (OSM). The steps to obtain and process the data are described in detail in \emph{Appendix A}. The data are structured as a multiplex network \citep{ battiston2014structural} with two different link types and three different node types, see Fig.~\ref{fig:gapdefinition}. Links of type ``unprotected'', shown in grey, denote street segments that are designed for motor vehicles and lack protected bicycle infrastructure. Links of type ``protected'', shown in black, denote protected bicycle infrastructure --- either alongside a street segment or off-street. If a node has only one type of links adjacent to it, we call the node either a protected node, shown in black, or an unprotected node, shown in grey. If a node has both protected and unprotected links adjacent to it, we call it a contact node, shown in blue.

We then define a gap as a shortest path between two contact nodes that consists only of unprotected links. This definition is based on the rationale that a gap should be a continuous piece of ``missing'' protected infrastructure, and it should be as short as possible. An example of a gap following this basic definition is illustrated in red in Fig.~\ref{fig:gapdefinition}.

For identifying gaps in our Copenhagen data set, we applied the Dijkstra all-pair shortest path algorithm to the entire street network with links weighted by length. From the set of paths obtained, we discarded all paths that do not meet our gap definition, i.e.~the start and end nodes must be contact nodes and all links must be unprotected. In this way, 9924 unique gaps were identified in our Copenhagen data set. 

\subsection*{Discarding parallel paths}
\label{sec:parallel_paths}
\begin{figure}[t!]
\centering
    \includegraphics[width=0.3\textwidth]{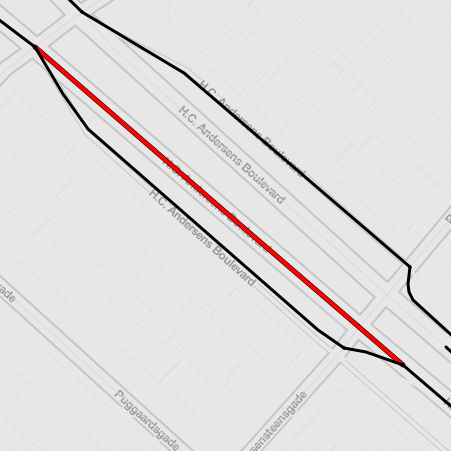}
    \includegraphics[width=0.3\textwidth]{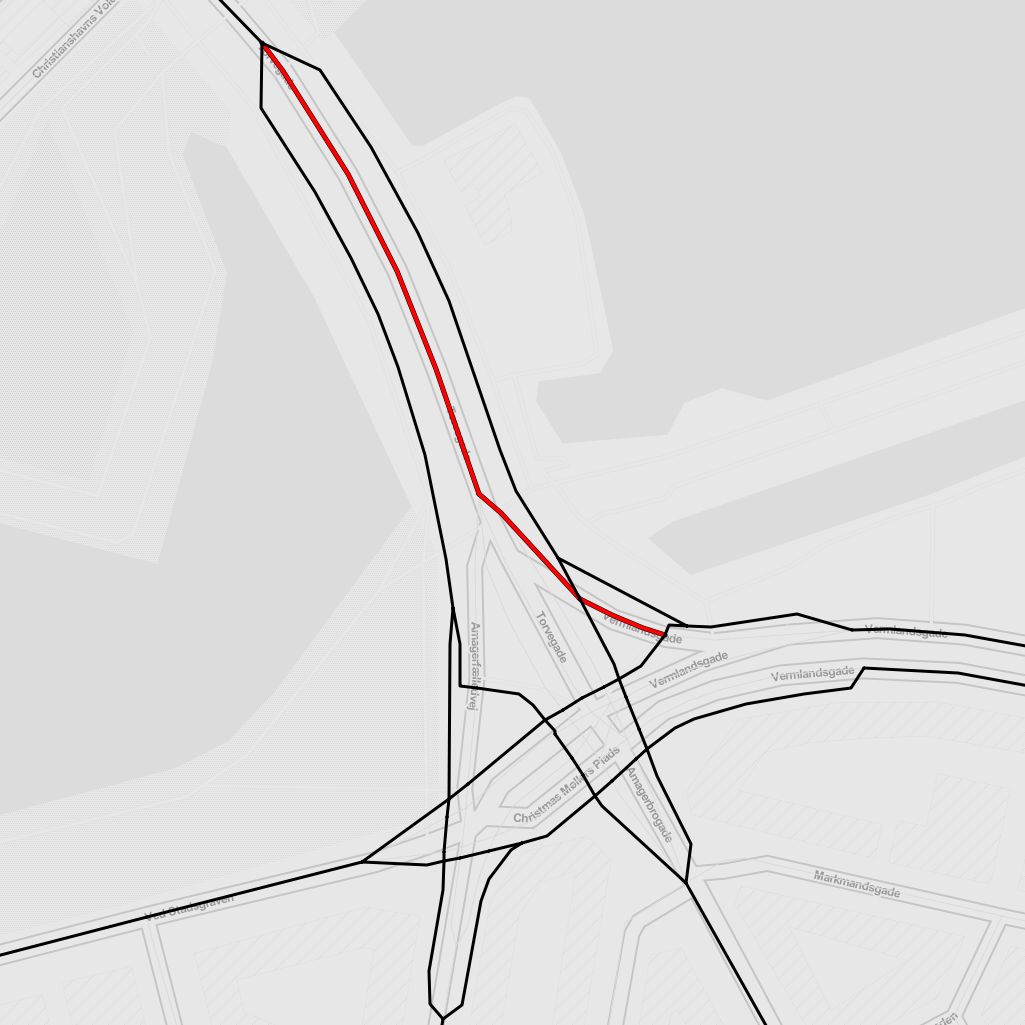}
    \caption{\textbf{Two examples of parallel paths that a gap identification process must account for.} The bicycle network is shown in black. The parallel paths along the car network are shown in red; they are only slightly longer and should not be identified as gaps. Left: Parallel path along H.C.~Andersens Boulevard. Right: Parallel path along Torvegade. \label{fig:pes}}
\end{figure}

Before proceeding to evaluating the benefits of closing gaps, we must ask whether our working definition of ``gaps'' will yield a meaningful set of potential ``missing links'', or whether we need to refine our approach. Indeed, applying our definition to Copenhagen's street network reveals the problem of \emph{parallel paths} that needs to be accounted for. This problem comes from the naive application of the shortest path algorithm which does not account for the common occurrence of protected off-street bicycle tracks that run in parallel to car lanes. In these cases, the shortest path algorithm with links weighted by length chooses the slightly shorter car path over the slightly longer bicycle path (see Fig.~\ref{fig:pes}), and therefore undesirably detects a gap located on the car lane despite a protected bicycle track running next to it. The parallel paths problem is a consequence of applying the shortest path algorithm to a relatively high-resolution network layer. However, lowering the resolution is not an option, because using map data with a high resolution of the street segments is necessary for identifying the gaps that we are looking for. This is a well-know problem in transportation network modeling: If a high-resolution layer is given as input, solving a routing problem at a lower resolution is a non-trivial task \citep{zhu_transportation_2015, perrine_map-matching_2015}. 

We therefore applied the following mitigation strategy for parallel paths: For each identified gap $\mathbf{g}$, we first computed the detour factor $D(\mathbf{g}) = \frac{d_{\textrm{prot}}(\mathbf{g})}{d_{\textrm{all}}(\mathbf{g})}$, where $d_{\textrm{prot}}(\mathbf{g})$ and $d_{\textrm{all}}(\mathbf{g})$ are the shortest network path distances on the network of protected bicycle infrastructure and on the entire street network, respectively. We then set a minimum detour value \(D_{\textrm{min}} = 1.5\) and discarded all previously identified gaps that had $D(\mathbf{g}) < D_{\textrm{min}}$.

We arrived at the detour factor value of 1.5 by manually comparing the results of applying a cut-off value for gap rank and the declustering heuristic (see sections \emph{Gap prioritization} and \emph{Gap declustering} below) first to the list of gaps with $D(\mathbf{g})\geq D_{\textrm{min}}$ and then to the list of gaps with $D(\mathbf{g})<D_{\textrm{min}}$, for different values of $1 < D_{\textrm{min}} < 2$. Setting $D_{\textrm{min}}=1.5$ yielded the fewest false positives and false negatives. For gaps with a detour factor of $D(\mathbf{g})\geq 1.5$, there were only 10\% of false positives, i.e.~gaps with a detour factor of over 1.5 that turned out to be parallel paths and had to be excluded manually. For the gaps with a detour factor $D(\mathbf{g})< 1.5$, we found three types of gaps: 1) an expected high percentage of 49\% of parallel edges, 2) in 43\% of cases a partial overlap with gaps of a higher detour factor and therefore no substantial loss of information when excluded, 3) only 8\% of false negatives, i.e.~actual gaps on the bicycle network. The chosen detour factor therefore presents a reasonable trade-off between minimizing false positives (roughly 10\% of the gaps that had to be excluded manually) and loss of information (roughly 8\% of automatically excluded gaps that were actually relevant). It is also in line with cyclist detour behavior reported in the literature \citep{reggiani_understanding_2021}.

Excluding gaps with a detour factor below 1.5 from our analysis reduces the number of gaps from 9924 to 6603. This list of 6603 identified gaps is used as input for the next step of the IPDC procedure: gap prioritization.

\subsection*{Gap prioritization}
Not all street segments that were identified so far as gaps are equally suitable for the construction of new bicycle infrastructure, nor are they equally relevant for the overall performance of the bicycle network. Consider and contrast two examples of locations without protected bicycle infrastructure: a  residential street in a suburban area, versus a narrow bridge over a canal in the city center. To put a number on the priority of each of these gaps, we need to ask not only ``How central is this missing link?'' and ``How much does it cost to close this gap?'', but also ``How many citizens will benefit from closing it?'' Therefore, after having found all gaps which fit our topological definition, the next step is to evaluate the benefits of ``closing a gap'' (by installing protected bicycle infrastructure) for the overall performance of the bicycle network, and to prioritize the list of gaps by this benefit metric.

To quantify the benefit of ``closing a gap'', we start off with the rationale that the positive impact consists in reducing the number of meters that cyclists have to ride in the same space as motorized traffic. This is in line with the concept of ``planning for the vulnerable'', i.e.~aiming to provide an inclusive transportation system by protecting the most vulnerable population groups --- such as children, who ideally should never have to cycle in mixed traffic \citep{noreen_mcdonald_children_2012}. If this concept was taken to the extreme, no single gap should be left unclosed, which is not a realistic goal. Therefore, we aim to approach this ideal \emph{most effectively} by prioritizing gaps that lie on the most commonly taken bicycle routes. Using topological street network data only, the most common routes can be gauged quantitatively by selecting gaps with the highest link betweenness centrality weighted by gap length. Let us provide an example before the formal definition. Assume that gap A has a length of $10\,\mathrm{m}$ and a traffic volume of 50 cyclists in a time unit (e.g.~during one hour); and gap B has a length of $20\,\mathrm{m}$ and a traffic volume of 15 cyclists. Then, by multiplying lengths with traffic volumes, we obtain the total number of meters cycled in mixed traffic: $500\,\mathrm{m}$ for gap A and $300\,\mathrm{m}$ for gap B. Closing gap A would avoid more meters cycled in mixed traffic, which is why gap A is ranked more relevant than gap B. In this case gap A is also shorter, therefore also more cost-efficient to close.

In order to apply this rationale, we estimated the number of cyclists on each link, i.e.~the bicycle traffic flow through the network, based on the network topology, using betweenness centrality. Betweenness centrality, derived from an all-pair shortest path algorithm, is the most basic proxy for traffic demand. It assumes that for each possible origin-destination combination, there is one ``cyclist unit'' making their way through the network, always choosing the shortest possible path between origin and destination. Then the number of cyclists that use a specific link on their way through the network, divided by the total number of cyclists on the network, will yield the fraction of cyclists that we expect to find on this link. Thus, the betweenness centrality indicates how ``central'' or relevant a link is for the flow of cyclists through the whole network. Similar approaches based on betweenness centrality have previously been used to estimate bicycle and motorized traffic flow \citep{mcdaniel_using_2014, jayasinghe_explaining_2015, ye_modified_2016}. This simple model can be refined arbitrarily by replacing betweenness with any other demand model, such as a gravity model, or with empirical flow data, but this refinement is outside the scope of this work.

\begin{figure}[t]
    \includegraphics[width=0.9\linewidth]{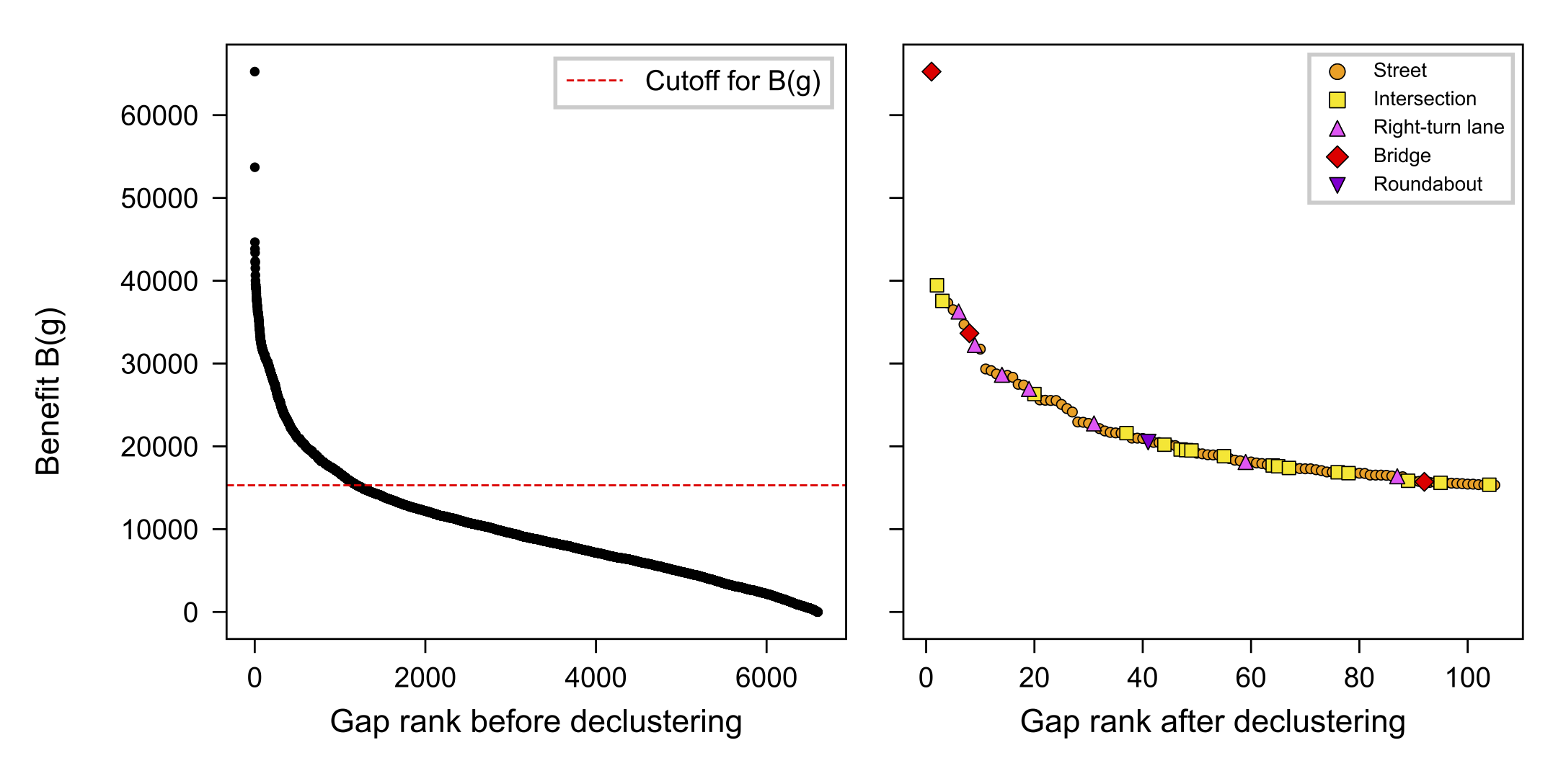}
    \caption{\textbf{Heterogeneity of gap closure benefits.} The distribution of the benefit metric $B(\mathbf{g})$ for the Copenhagen gaps before (left) and after (right) prioritization, declustering and error correction. The highest ranked gaps have a much higher benefit for the overall network; bridges tendentially fall into this category. The dashed red line on the left plot shows the benefit cut-off value used in the prioritization step. The colors in the left plot represent different gap classes (see section \emph{Gap classification}).}
    \label{fig:ranking_param}
\end{figure}

There is a non-trivial dependence of flow-based centrality metrics on changes in network boundaries. This phenomenon, known as ``network edge effect'' \citep{okabe_spatial_2012} or ``border effect'' \citep{porta_network_2006}, also has relevant implications for equity considerations, since centrality metrics like betweenness have an inherent bias towards the center of the network. To account for this network edge effect, we introduced a cut-off radius $\lambda$ for the set of shortest paths, based on which the centrality metrics are computed \citep{gil_street_2017, yamaoka_local_2021}. Setting this locality parameter $\lambda = \infty$ in the shortest path algorithm would consider the entire street network for origin-destination pairs, finding gaps that are most relevant for the whole city and have a tendency to be located more centrally. By contrast, we used $\lambda=2500\,$m to include only destination nodes that are within a maximum path length of $2500$ meters from each origin node, finding gaps that are relevant for sub-city scale flows, e.g.~on district or neighborhood scales. We chose this particular length as it roughly corresponds to the average diameter of the administrative districts of Copenhagen \citep{trap_danmark_kobenhavns_2021}, and lies in the range of 2--4.9$\,$km, which is the most frequent bicycle trip length range within Copenhagen \citep{kobenhavns_kommune_teknik-_og_miljoforvaltningen_fra_2021}. By using a finite $\lambda$ in our calculations, we obtain several benefits: the bias towards the center of the network is decreased; the local importance of the identified gaps can be regulated; and, lastly, computation time is substantially reduced, which is particularly relevant for larger cities.

We applied the locality parameter $\lambda$ to the set of all shortest paths $\mathcal{P}$ to compute the link betweenness centrality ${c}_{\lambda}(l) = \sum_{d(i,j)<\lambda} n_{l}(i,j)$ for each link $l$, where $n_{l}(i,j)$ is the number of times the link $l$ appears in $\mathcal{P}$. By multiplying the betweenness centrality ${c}_{\lambda}(l)$ of link $l$ by its length $L(l)$, we obtained the total number of expected meters cycled on this link, the link closure benefit $B^*_\lambda(l)$. Since a gap $\mathbf{g}$ can consist of several links, the gap closure benefit $B^*_\lambda(\mathbf{g})$ is obtained from adding up the link closure benefits of each of the links $l$:

\begin{equation}
B^*_\lambda(\mathbf{g}) = \sum_{l\,\in\,\mathbf{g}} {c}_{\lambda}(l) \cdot L(l)
\end{equation}

\noindent As a last step, we account for cost-efficiency. We assume for simplicity, and in line with previous studies \citep{mauttone_bicycle_2017}, that construction costs are generally proportional to facility length. We therefore divide the expected meters cycled $B^*_\lambda(\mathbf{g})$ by the gap's total length $L(\mathbf{g}) = \sum_{l \in \mathbf{g}} L(l)$ and thus obtain the expected meters cycled \emph{per investment unit} that would be avoided if the gap was closed:
\begin{equation}
B_\lambda(\mathbf{g}) = \frac{B^*_\lambda(\mathbf{g})}{L(\mathbf{g})} \label{eq:benefit}
\end{equation}
\noindent This model is extendable with further weights, for example with data on specific road hazards or stress levels \citep{furth_network_2016, chen_how_2017}, or by a non-linear cost function. However, for sake of simplicity and generality, we do not assign any further weights here. This  corresponds to the simplifying assumption that for each cyclist, every meter cycled jointly with motorized traffic equally contributes to the risk of getting injured or killed. Note that for the case of gaps that consist of only one link, eq.~(\ref{eq:benefit}) simplifies to $B_\lambda(\mathbf{g})={c}_{\lambda}(l)$ since the two expressions for total gap length cancel out. From here onwards, we drop the index $\lambda$ for simplicity and denote the gap closure benefit as $B(\mathbf{g})$.

The benefit metric $B(\mathbf{g})$ will be used for gap prioritization. To summarize, it expresses the benefits of closing a gap $\mathbf{g}$ in terms of number of expected meters cycled in mixed traffic per unit of investment. Fig.~\ref{fig:ranking_param} (left) shows the distribution of $B(\mathbf{g})$ for Copenhagen within the list of 6603 gaps that were found with a minimum detour of $D_{\textrm{min}}\geq 1.5$. This distribution shows a large heterogeneity of benefits due to the heavy-tailed distribution of betweenness in street networks \citep{kirkley_betweenness_2018}. In other words, there is a small subset of highest-ranked gaps which account for a substantial amount of the total benefit. After inspecting the heterogeneity of benefits, Fig.~\ref{fig:ranking_param} (left), we chose a cut-off at $B(\mathbf{g}) \geq 15\,000$ where the growth of the rank-ordered benefits changes qualitatively, thereby selecting approximately the highest ranked 20\% of gaps. This selection provides an ordered list of the 1199 highest-ranked gaps which is used as input to the next step of our IPDC procedure: gap declustering.

\subsection*{Gap declustering}

\begin{figure}[t]
    \centering
    \begin{subfigure}{.34\linewidth}
    \includegraphics[width=\linewidth]{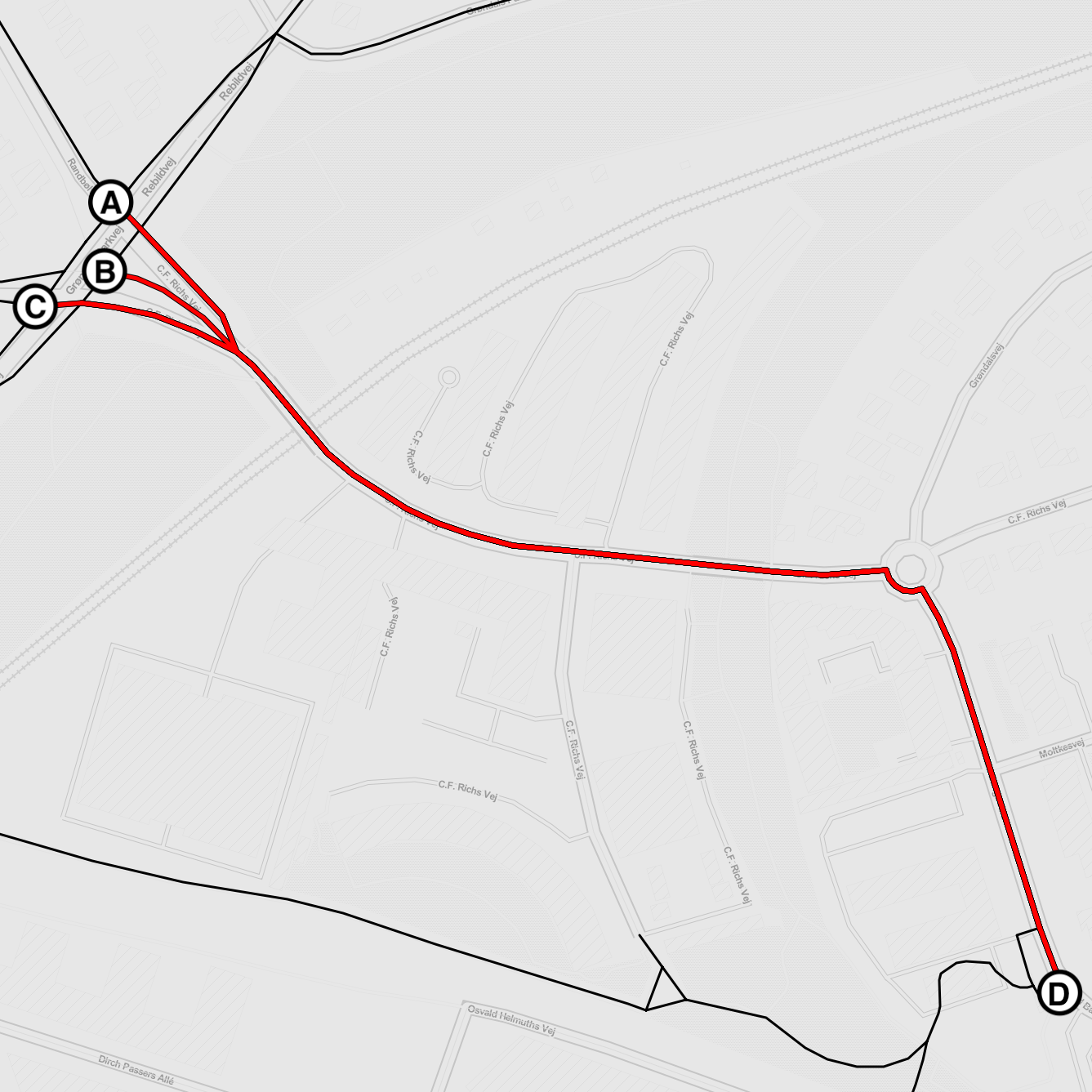}
    \end{subfigure}\hspace{0.75cm}
    \begin{subfigure}{.34\linewidth}
    \includegraphics[width=\linewidth]{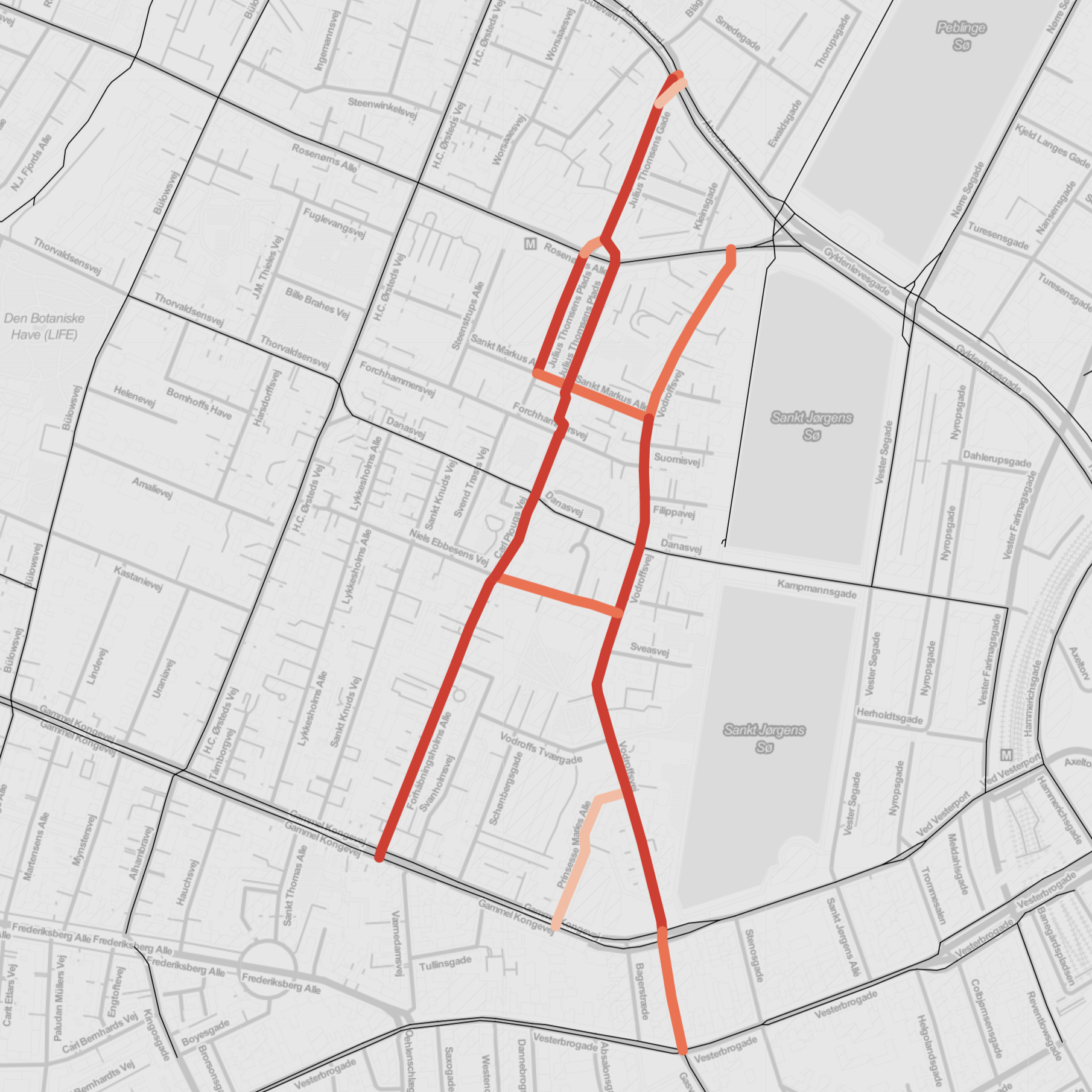} 
    \end{subfigure}
    \begin{subfigure}{.169\linewidth}
    \includegraphics[width=\linewidth]{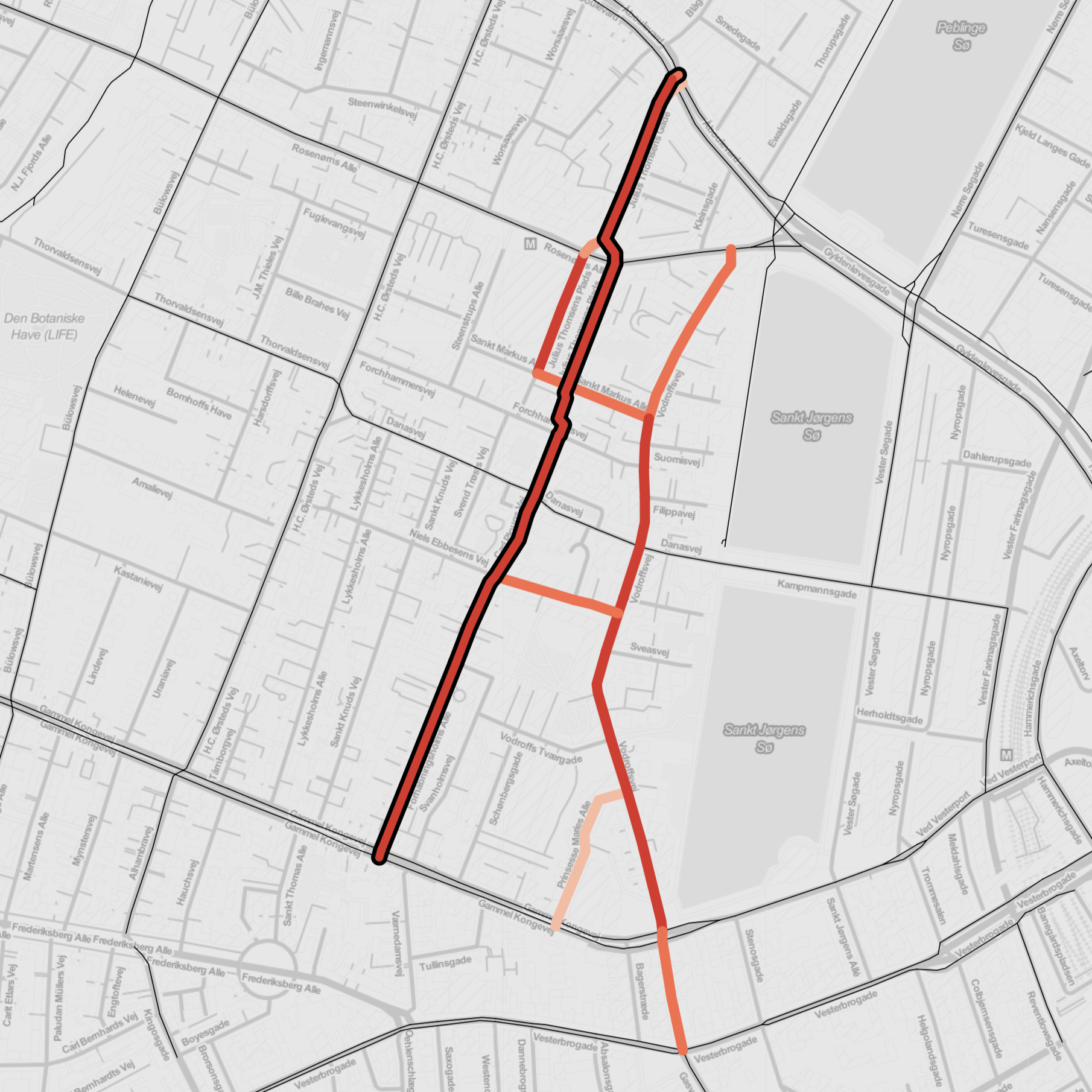}
    \includegraphics[width=\linewidth]{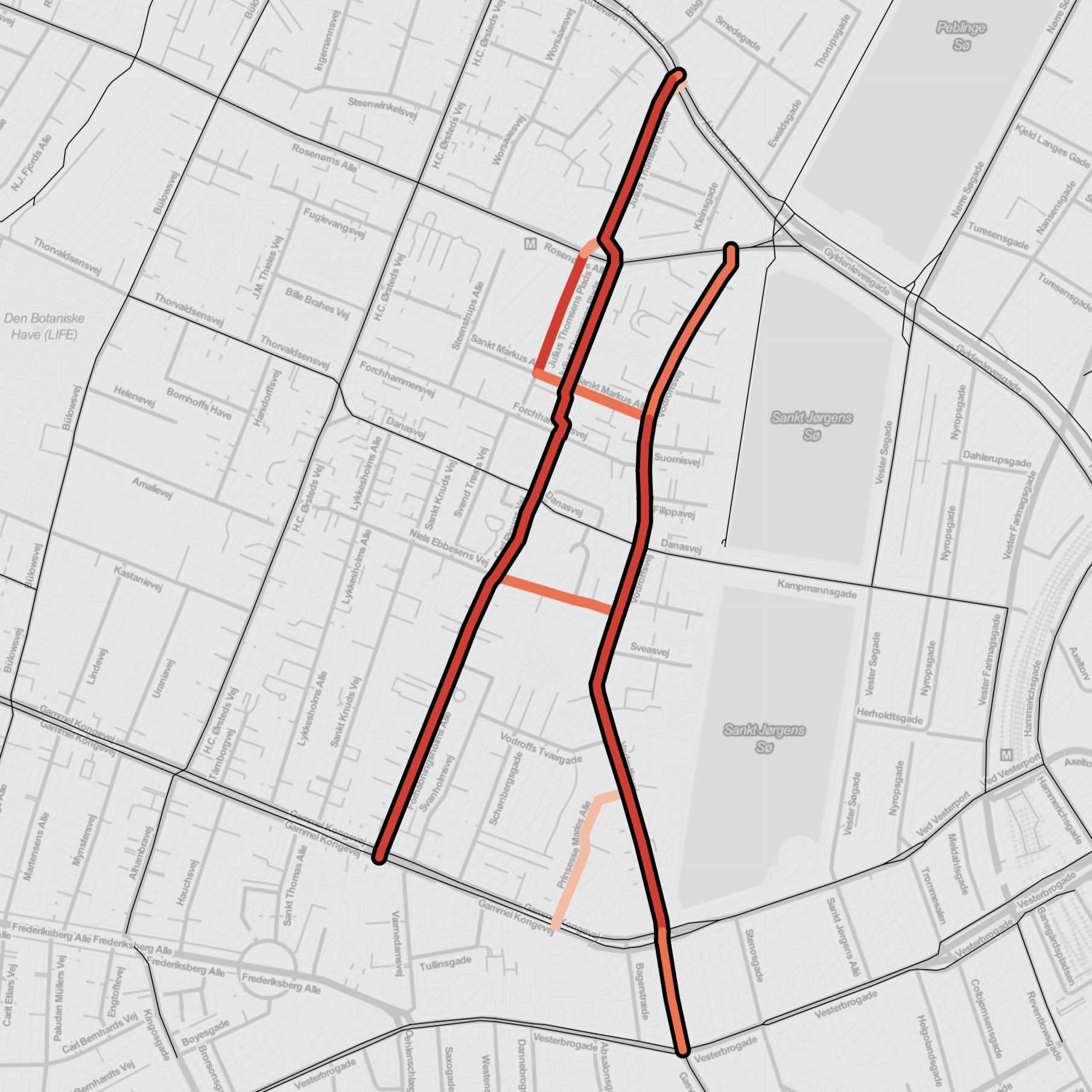} 
    \end{subfigure}
    \caption{\textbf{Example of gap clusters, and of the declustering heuristic.} Left: Shown in red, on C.F.~Richs Vej. The three gaps AD, BD and CD overlap and have similar closure benefits $B(\mathbf{g})$. Therefore, the three gaps are merged into one gap cluster to be handled jointly. Right: A declustering heuristic can help deciding which parts to retain and which ones to discard. The links of the shown gap cluster are colored by betweenness centrality; darker tones represent higher values. The black borders indicate the declustered gaps after one and two runs of the heuristic.\label{fig:fx_cluster}}
\end{figure}

In many cases, two or more prioritized gaps partially overlap, forming a \emph{gap cluster} that is not a simple path anymore. An example of gap clustering in Copenhagen is shown in Fig.~\ref{fig:fx_cluster} (left): Because the intersection of C.F.~Richs Vej and Grøndals Parkvej is represented by several network nodes, all shortest paths to destination node D from any of the origin nodes A, B, C are classified as gaps and display similar benefit values $B(\mathbf{g})$. This gap cluster example illustrates that it is not always meaningful to provide all street segments that constitute a gap cluster with protected bicycle infrastructure. The appearance of gap clusters in the results can also be understood by recalling our gap-finding procedure and network characteristics: First, all network are considered as equally likely origins or destinations; second, gaps consist of car links and start and end on a contact node; and third, the network is characterized by a high node density e.g.~at intersections of streets with multiple lanes. Taken together, these three points help explain that gaps will often consist of a combination of high and low centrality links, and moreover, gaps will often partially overlap, meaning that the same street segment will appear in several gaps --- for example, the network link on Brønshøjvej appears in more than 100 of the 6603 gaps found. 

The urban planning task of identifying the exact subnetworks within these gap clusters for the construction of infrastructure to ``close the gap'' is beyond the scope of the present study. However, we developed a declustering heuristic, described in detail in \emph{Appendix A}, which is a first approach to break down a gap cluster into separate components that are simple paths, based on the same benefit metric derived $B(\mathbf{g})$ from betweenness centrality. See Fig.~\ref{fig:fx_cluster} (right) for a non-trivial gap cluster with links colored by edge betweenness centrality values and the resulting declustered gaps. After applying the gap declustering heuristic, using the list of 1199 highest-ranked gaps as input, we obtain a list of 134 gaps. This list is used as input for the next step of the IPDC procedure: gap classification.

\subsection*{Gap classification}
The last step in the IPDC procedure is the classification of gaps. The gap classification scheme described in this section was developed through manual inspection and on-site visits of the gaps identified in Copenhagen, hence it might need to be adapted or extended for other urban contexts in future research. We identified the following gap classes: Street (ST); intersection (IS); right-turn lane (RT); bridge (BR); roundabout (RA); and error (ER). This classification scheme is meant to facilitate both the interpretation of results from bicycle network analysis and the decision-making within a subsequent planning process. In this section, we describe the general concept behind each of the gap classes before discussing the specific results for Copenhagen.

\paragraph{Street} The gap class \emph{street} corresponds most intuitively to the idea of a ``gap in the bicycle network'', i.e.~a generic street segment without protected bicycle infrastructure. We define as street gap all mixed-traffic street segments whose both ends connect to protected bicycle infrastructure and that do not correspond to any of the other gap classes (bridge, intersection, roundabout, right-turn lane, or error). 

\paragraph{Intersection} 
Missing links without protected bicycle infrastructure found at crossings of two or more streets are classified as \emph{intersection}. Given that a high proportion of traffic crashes occur at intersections, intersection design is crucial for cyclist safety \citep{thomas_safety_2013}. By the very nature of an intersection, a potential for conflict between traffic participants cannot be brought to zero; however, it can be minimized with appropriate planning \citep{crow2016dmb}. Intersection design deserves to be considered a discipline of its own right, and different network analysis methods than the one used in this study might need to be applied to explicitly identify problematic intersections from a bicycle network planning perspective \citep{furth_network_2016}. 

In the present study, we do not model intersections separately, but rather identify them as gap class in the last step of the procedure. Due to the underlying data structure in OSM, intersections could only be identified as gaps within the IPDC procedure if they contained at least one link, rather than just nodes. Additionally, there is a lack of consistency in OSM tagging when it comes to the designation of specific intersection segments as ``protected'' or ``unprotected''. The caveats of this approach are addressed in more detail in Section \emph{Scope and limitations} and in \emph{Appendix A}. 

\begin{figure}[t]
\captionsetup[subfigure]{labelformat=empty}
    \begin{subfigure}[t]{0.19\linewidth}
    \centering
    \includegraphics[width=\textwidth]{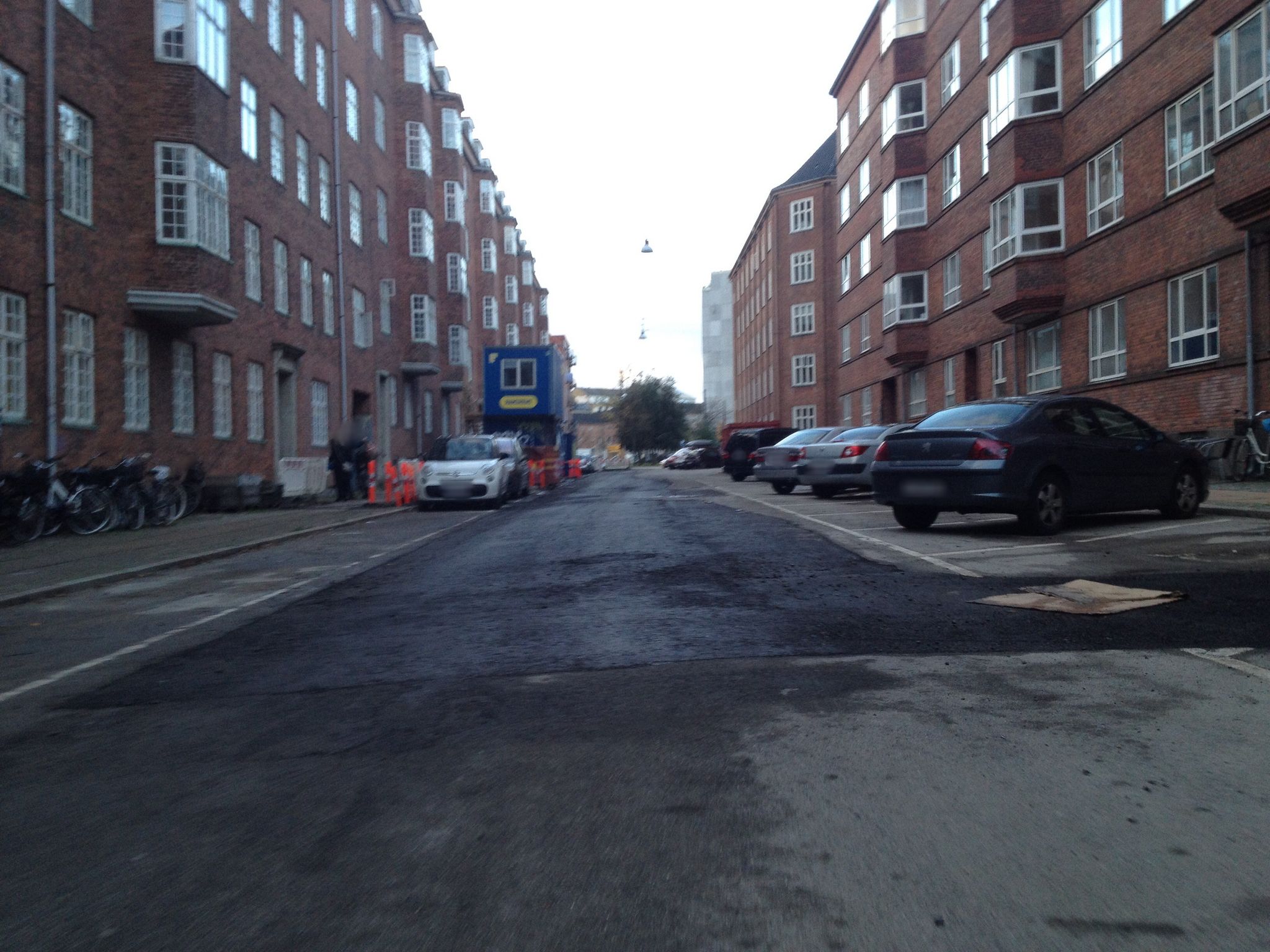}
    \end{subfigure}
    \begin{subfigure}[t]{0.19\linewidth}
    \centering
    \includegraphics[width=\textwidth]{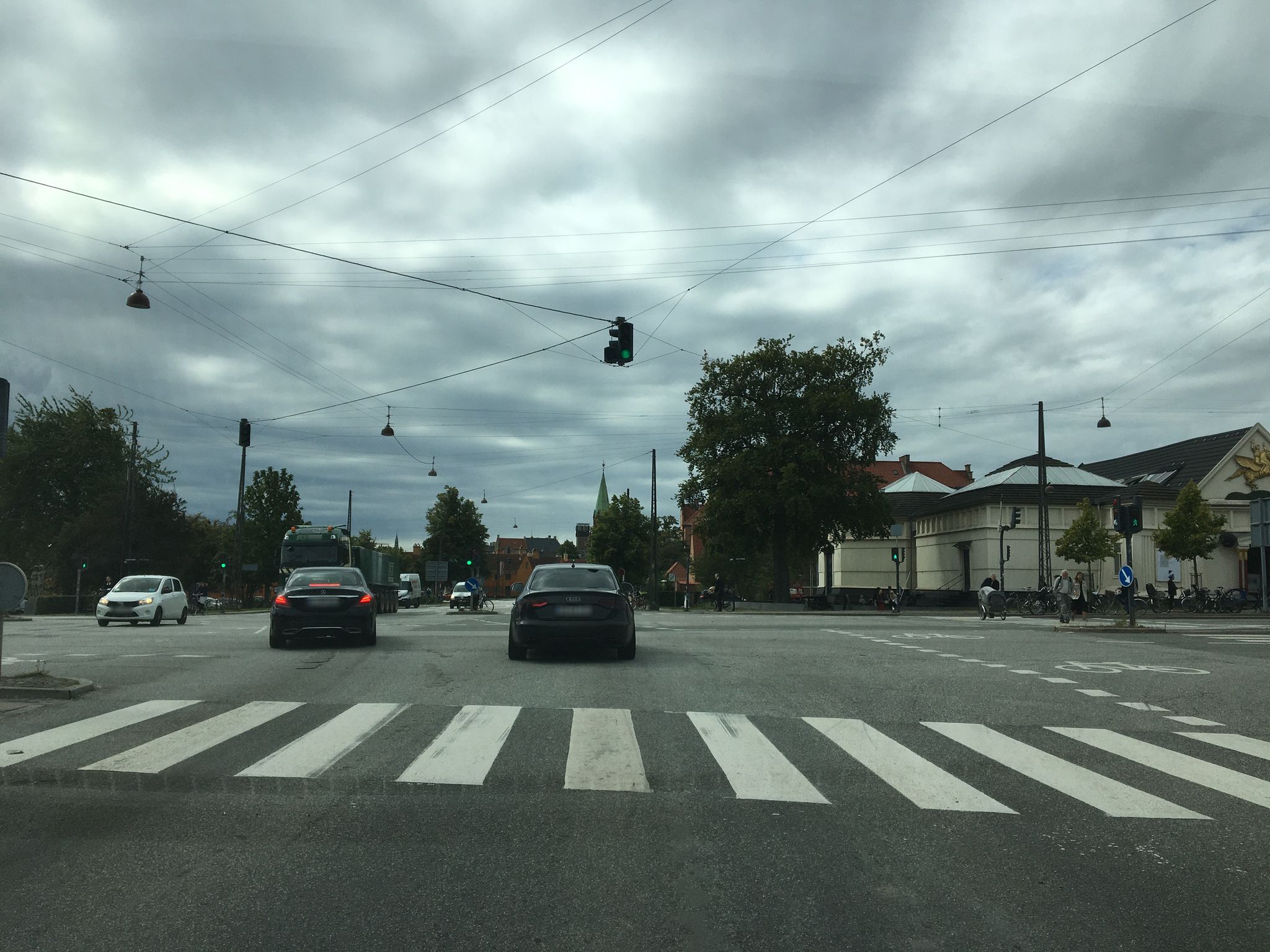}
    \end{subfigure}
    \begin{subfigure}[t]{0.19\linewidth}
    \centering
    \includegraphics[width=\textwidth]{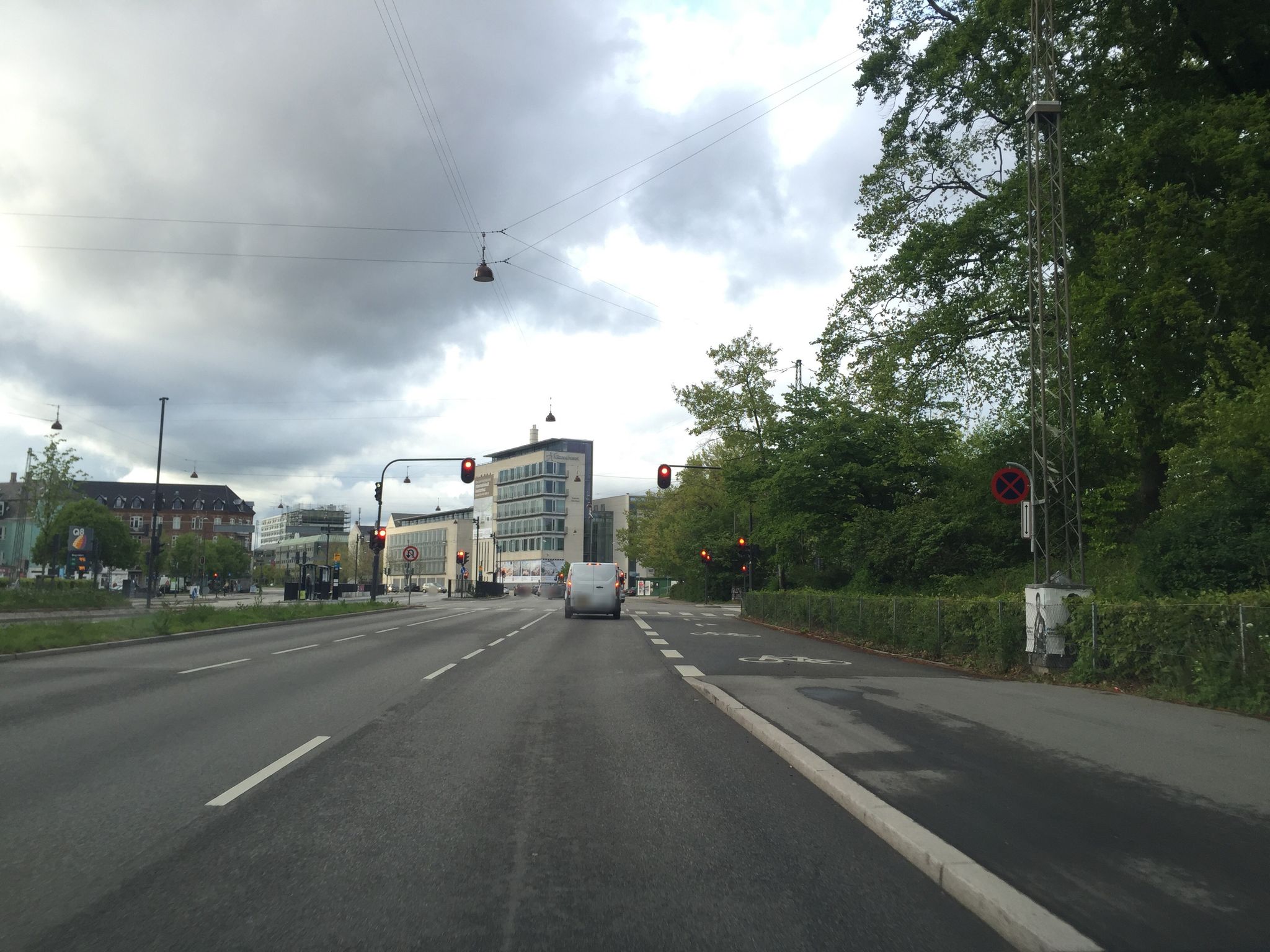}
    \end{subfigure}
    \begin{subfigure}[t]{0.19\linewidth}
    \centering
    \includegraphics[width=\textwidth]{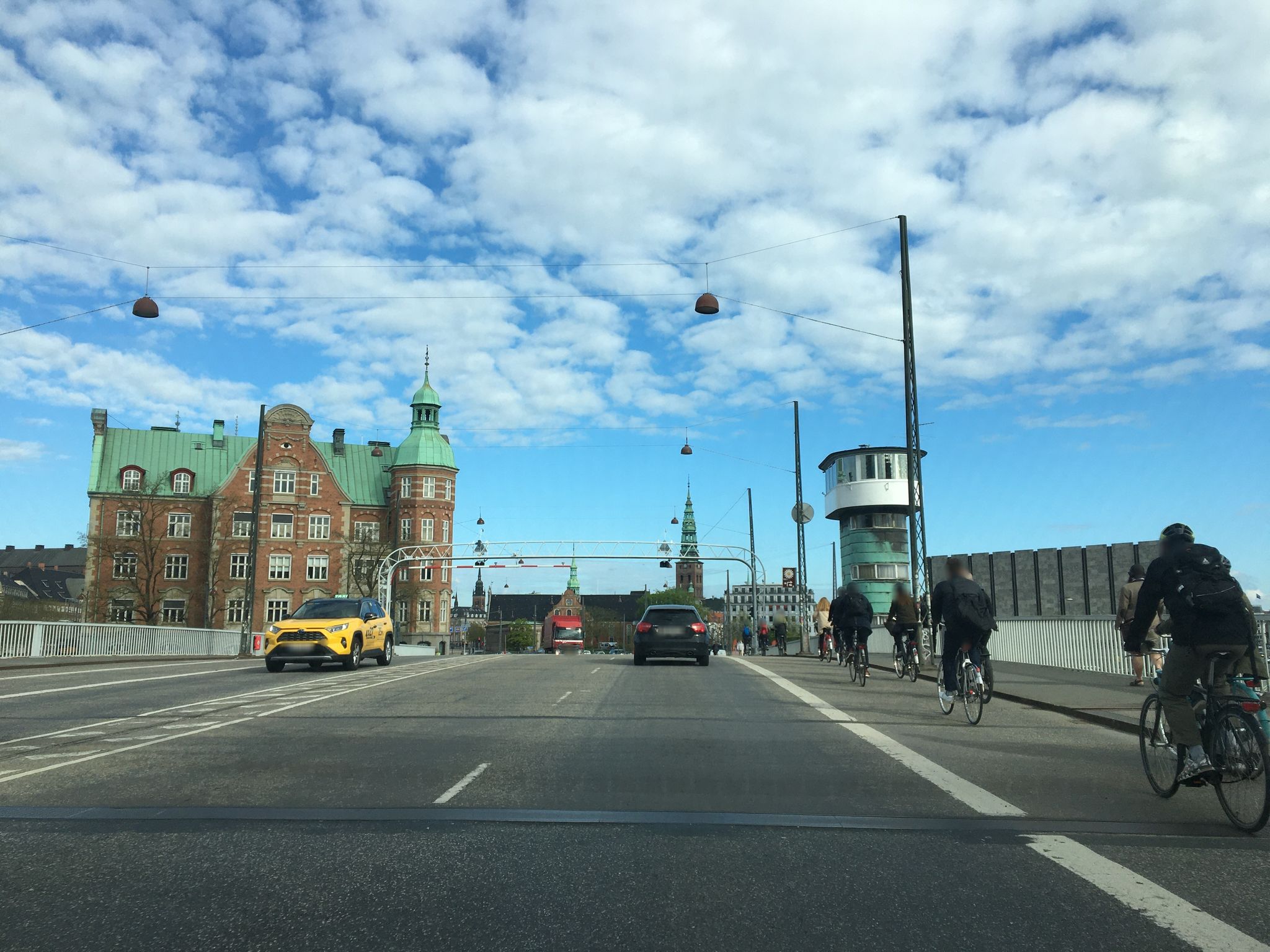}
    \end{subfigure}
    \begin{subfigure}[t]{0.19\linewidth}
    \centering
    \includegraphics[width=\textwidth]{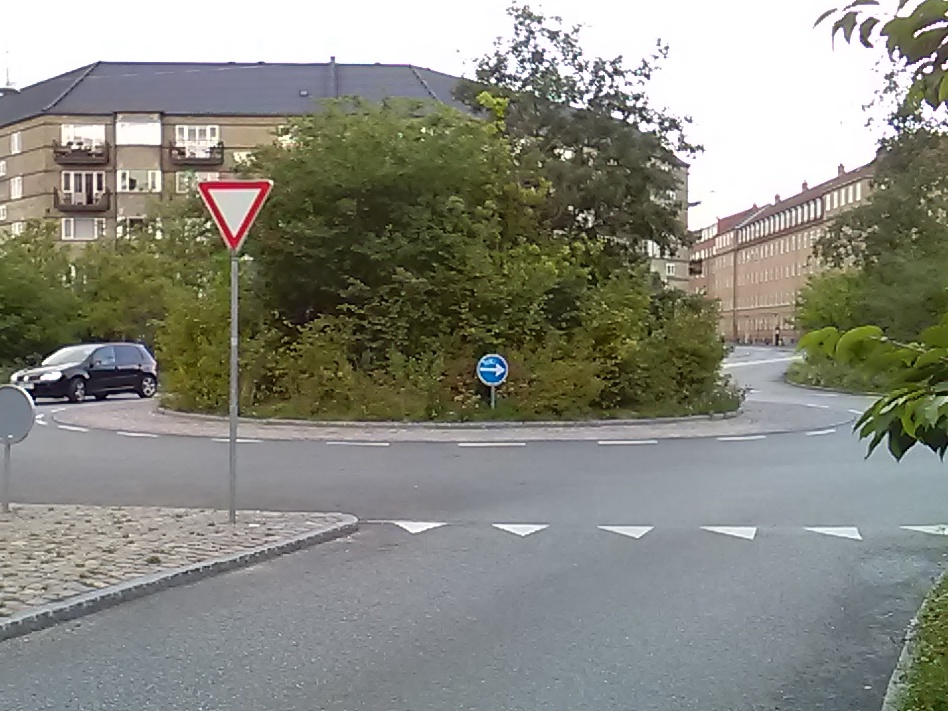}
    \end{subfigure}
\\
    \begin{subfigure}[t]{0.19\linewidth}
    \centering
    \includegraphics[width=\textwidth]{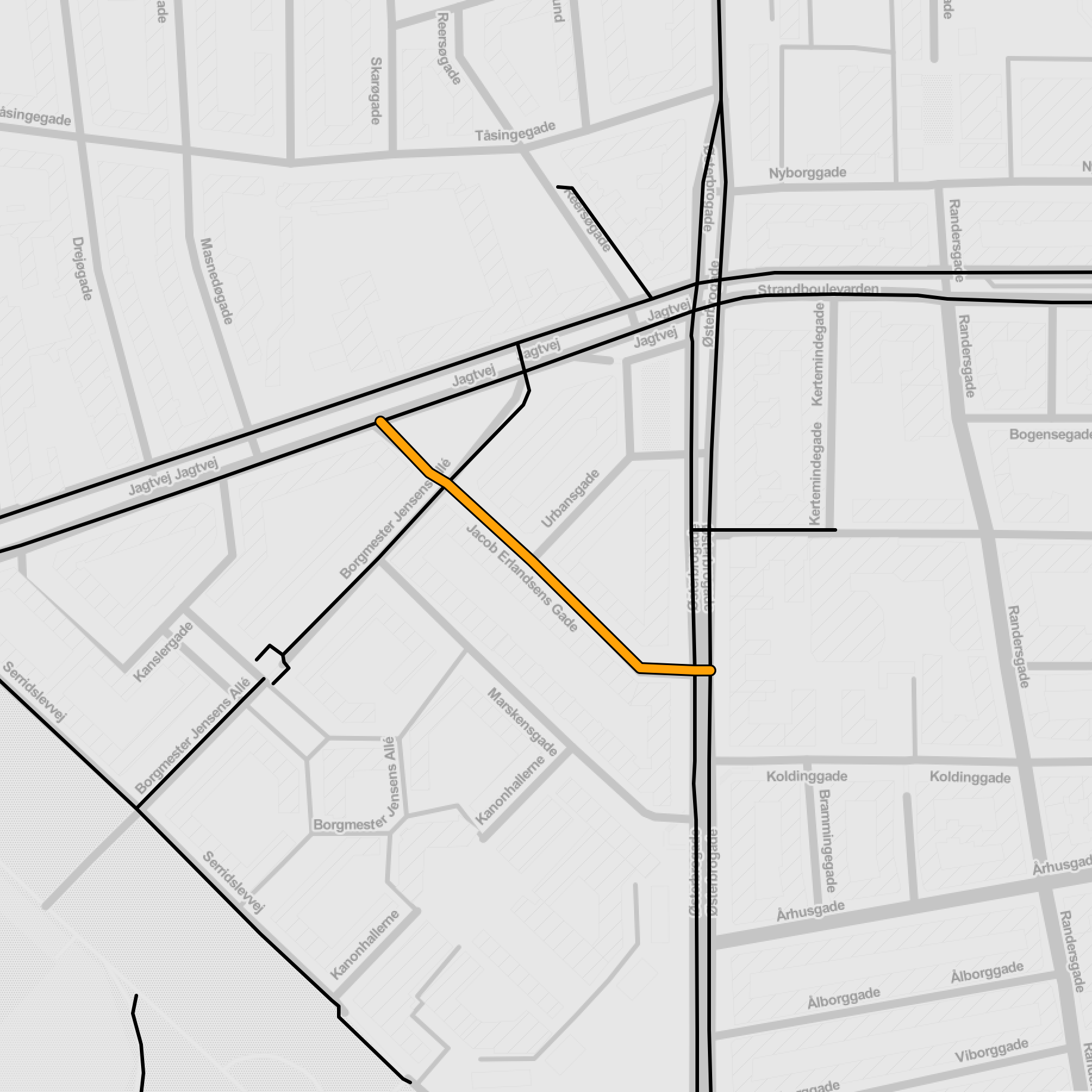} 
    \caption{Street}
    \end{subfigure}
    \begin{subfigure}[t]{0.19\linewidth}
    \centering
    \includegraphics[width=\textwidth]{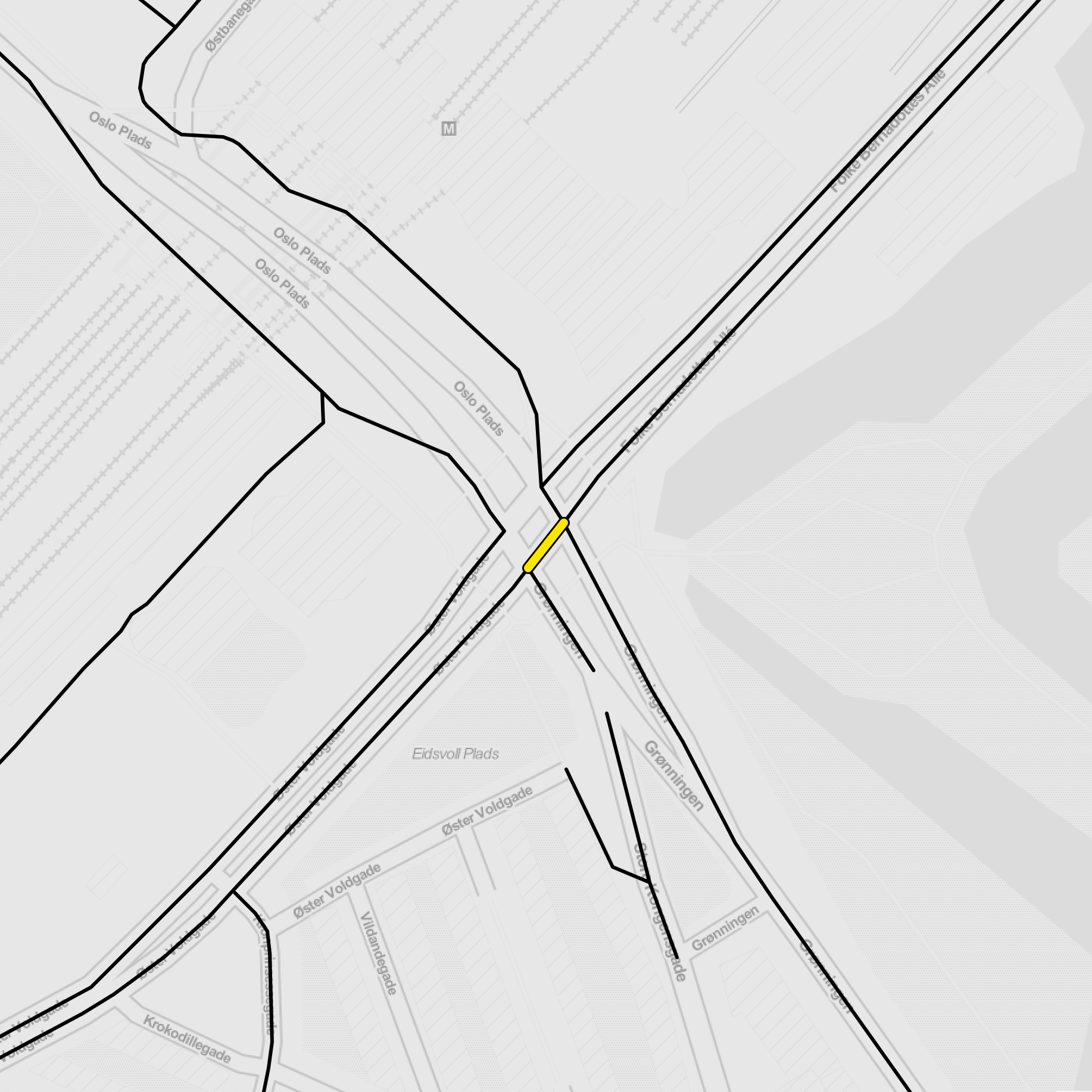} 
    \caption{Intersection}
    \end{subfigure}
    \begin{subfigure}[t]{0.19\linewidth}
    \centering
    \includegraphics[width=\textwidth]{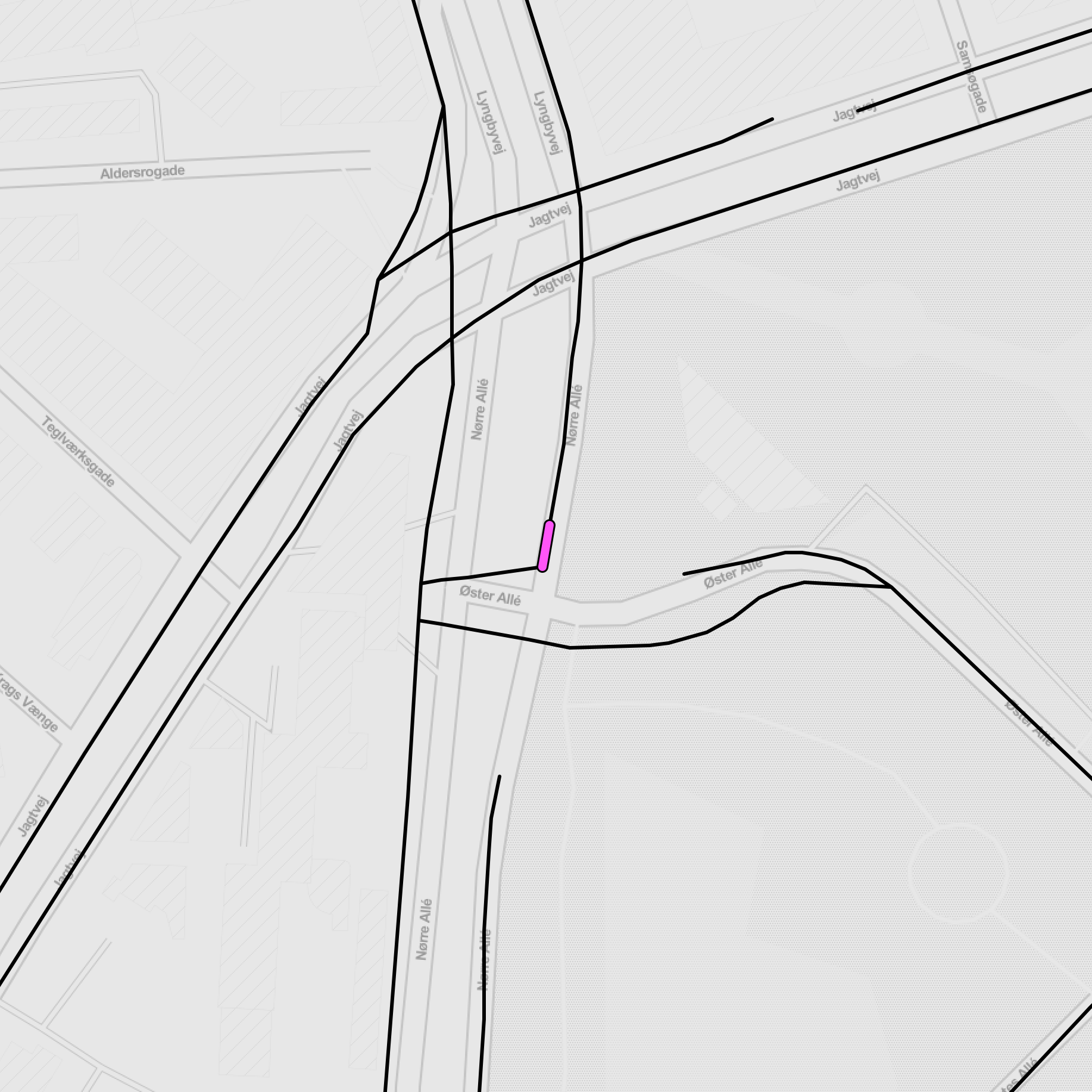} 
    \caption{Right-turn lane}
    \end{subfigure}
    \begin{subfigure}[t]{0.19\linewidth}
    \centering
    \includegraphics[width=\textwidth]{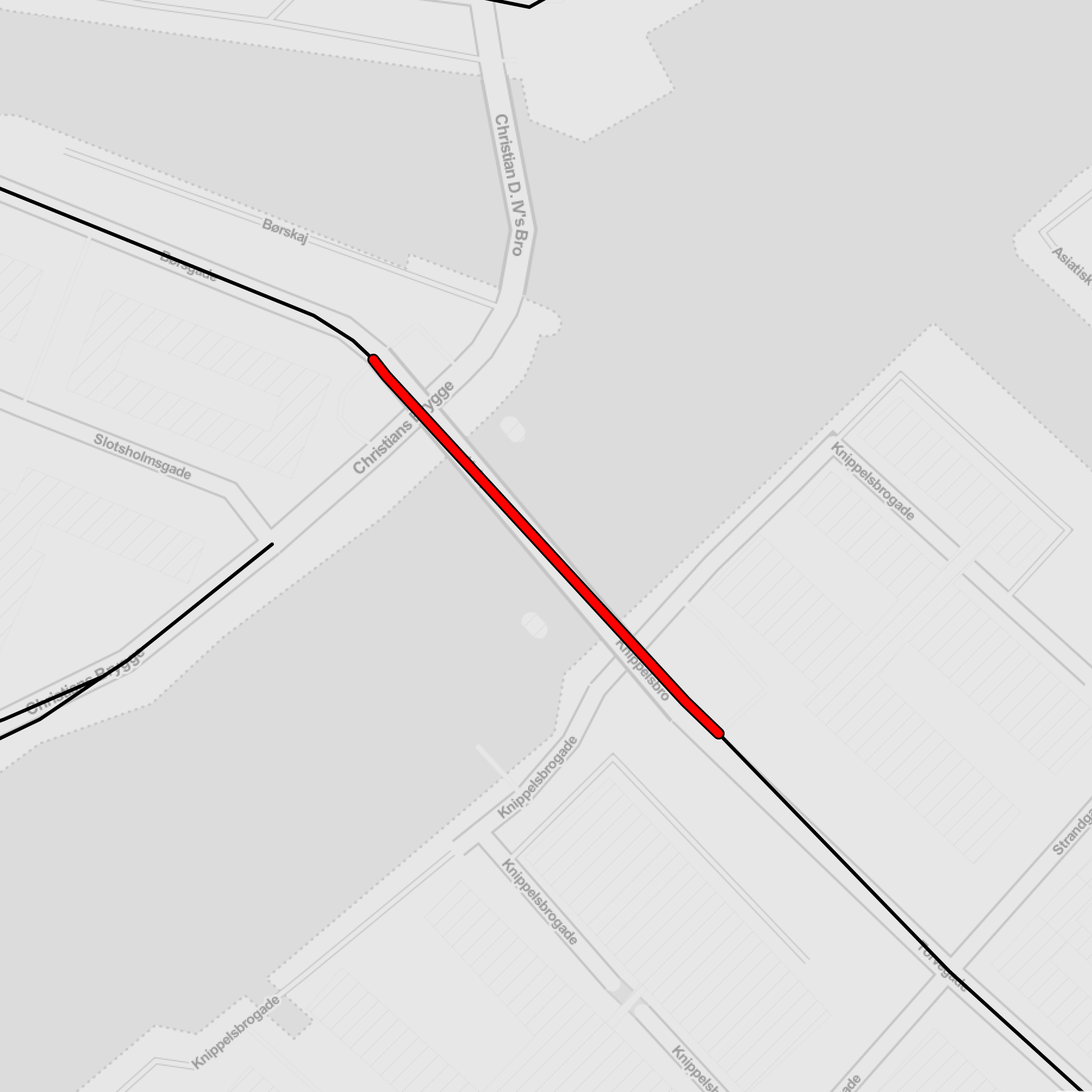} 
    \caption{Bridge}
    \end{subfigure}
    \begin{subfigure}[t]{0.19\linewidth}
    \centering
    \includegraphics[width=\textwidth]{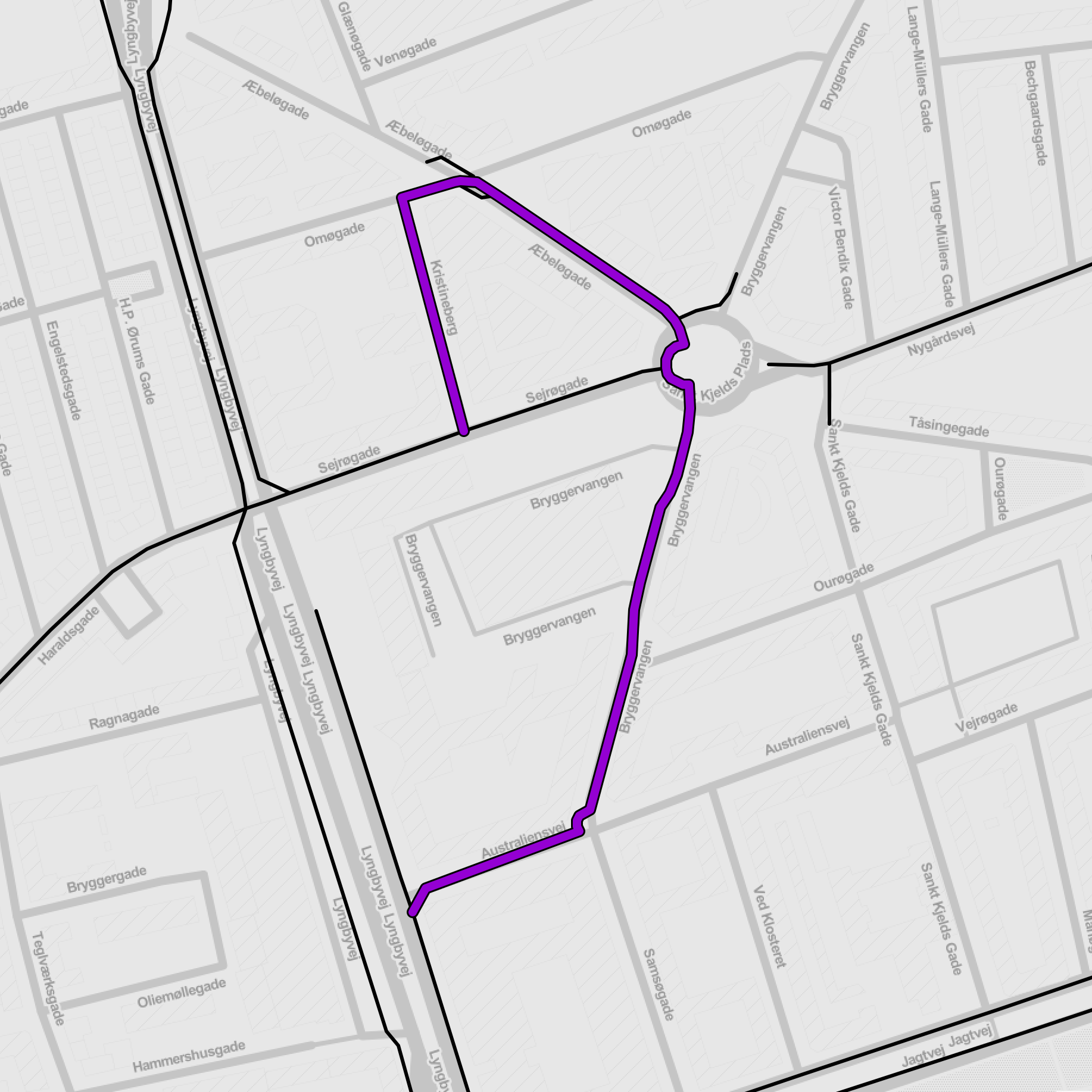} 
    \caption{Roundabout}
    \end{subfigure}
\caption{\textbf{The five gap classes in Copenhagen.} For each class, the highest ranked gap is shown. From left to right: Street gap (rank 4) on Jacob Erlandsens Gade; Intersection gap (rank 2) at the intersection of Øster Vøldgade and Grønningen; Right-turn lane gap (rank 6) at the right-turn from Nørre Allé to Øster Allé; Bridge gap (rank 1) on Knippelsbro; Roundabout gap (rank 41) at Sankt Kjelds Plads.}
    \label{fig:5gapclasses}
\end{figure}

\paragraph{Right-turn lane} 
We classify intersection approaches where the lane for right-turning cars merges with the adjacent cycle lane as \emph{right-turn lane} gaps. In such cases, the bicycle path ceases to be part of the protected bicycle network as it approaches an intersection, and cyclists are forced to mix with motorized traffic --- see Fig.~\ref{fig:5gapclasses} for an example. This type of intersection approach design is a common feature of Copenhagen's bicycle network \citep{vejdirektoratet_prevent_2017}. The Danish Road Directorate argues in favour of an intersection approach design with shortened cycle tracks where cyclists and cars mix for right turns \citep{sorensen_evaluering_2020, vejdirektoratet_vejtekniske_2020}, while current international best practice standards recommend intersections that protect and prioritize cyclists \citep{wagenbuur_junction_2014, crow2016dmb, national_association_of_city_transportation_officials_nacto_dont_2019}, such as dedicated bicycle queue areas and corner wedges or islands. Here we adhere to the international standards and to the rationale of demanding continuity for the network of protected bicycle infrastructure. 

\paragraph{Bridge} 
We classify missing links on obstacle-crossing road segments as \emph{bridge} gaps. In locations where there are physical barriers such as water bodies or railway tracks that have to be crossed, bridges play a particularly important function for connecting parts of the network and often constitute bottlenecks for traffic flow. At the same time, there are often inherent constraints to placing additional infrastructural elements on bridges due to limited physical space available \citep{wang_spatio-temporal_2019}. 

\paragraph{Roundabout} 
Since requirements for roundabout design are not the same as for intersections, we separately define the gap class \emph{roundabout}. Roundabouts are often considered to be the safer option for cyclists \citep{ dufour_presto_2010, jensen_safe_2017, us_department_of_transportation_federal_highway_administration_proven_2017}, depending on traffic volume \citep{crow2016dmb}. A roundabout with more than one lane puts cyclists at danger \citep{dufour_presto_2010}. According to a recent literature review by \cite{poudel_bicycle_2021}, data from Northern Europe suggests that the number of bicycle crashes might actually be higher for roundabouts than for intersections. There are several roundabout design options focusing on cyclist safety \citep{sakshaug_cyclists_2010}, such as the Zwolle roundabout, named after the Dutch city that first introduced it \citep{crow2016dmb, wagenbuur_experimental_2013}. 

\paragraph{Error} 
We classify gaps that have been identified by the IPDC procedure, but were not confirmed as such via visual inspection, as \emph{errors}. There are two types of errors: \emph{parallel paths} and \emph{data issues}. Parallel paths, as described in section \emph{Discarding parallel paths} above, are errors stemming from the routing problem in high resolution networks. Data issues are errors due to incorrect information on OSM. There are many possible reasons for errors in the OSM data: segments might be missing, mistagged, or outdated. The implications of OSM data quality on the results of this study are discussed in detail in section \emph{Scope and limitations}.

\begin{table}[t]
    \centering
    \begin{tabular}{|c|c|c|c|c|c|}
    \hline
    \textbf{Color} & \textbf{Acronym} & \textbf{Gap type} & \textbf{Count} & \textbf{Average benefit $\langle B \rangle_\mathbf{g}$} \\
    \hline
        \tikz{\filldraw[color=black, fill=myorange, very thick](0,0) rectangle (0.3cm,0.3cm);} & \textbf{ST} & Street & 77 &  $20\,544$ \\
        \tikz{\filldraw[color=black, fill=myyellow, very thick](0,0) rectangle (0.3cm,0.3cm);} & \textbf{IS} & Intersection & 17 & $20\,925$ \\
        \tikz{\filldraw[color=black, fill=mypink, very thick](0,0) rectangle (0.3cm,0.3cm);} & \textbf{RT} & Right-turn lane & 7 & $25\,911$ \\
        \tikz{\filldraw[color=black, fill=myred, very thick](0,0) rectangle (0.3cm,0.3cm);} & \textbf{BR} & Bridge & 3 &  $38\,207$ \\
        \tikz{\filldraw[color=black, fill=myviolet, very thick](0,0) rectangle (0.3cm,0.3cm);} & \textbf{RA} & Roundabout & 1 & $20\,518$ \\
    \hline
    \end{tabular}
    \caption{\textbf{Distribution of gap classes for the top 105 gaps in Copenhagen.} Bridges are most important.
    \label{tab:classes_dist}}
\end{table}

\section*{Finding gaps in Copenhagen with IPDC: The top 105 gaps}
\begin{figure}[p]
    \centering
    \includegraphics[width=1\textwidth]{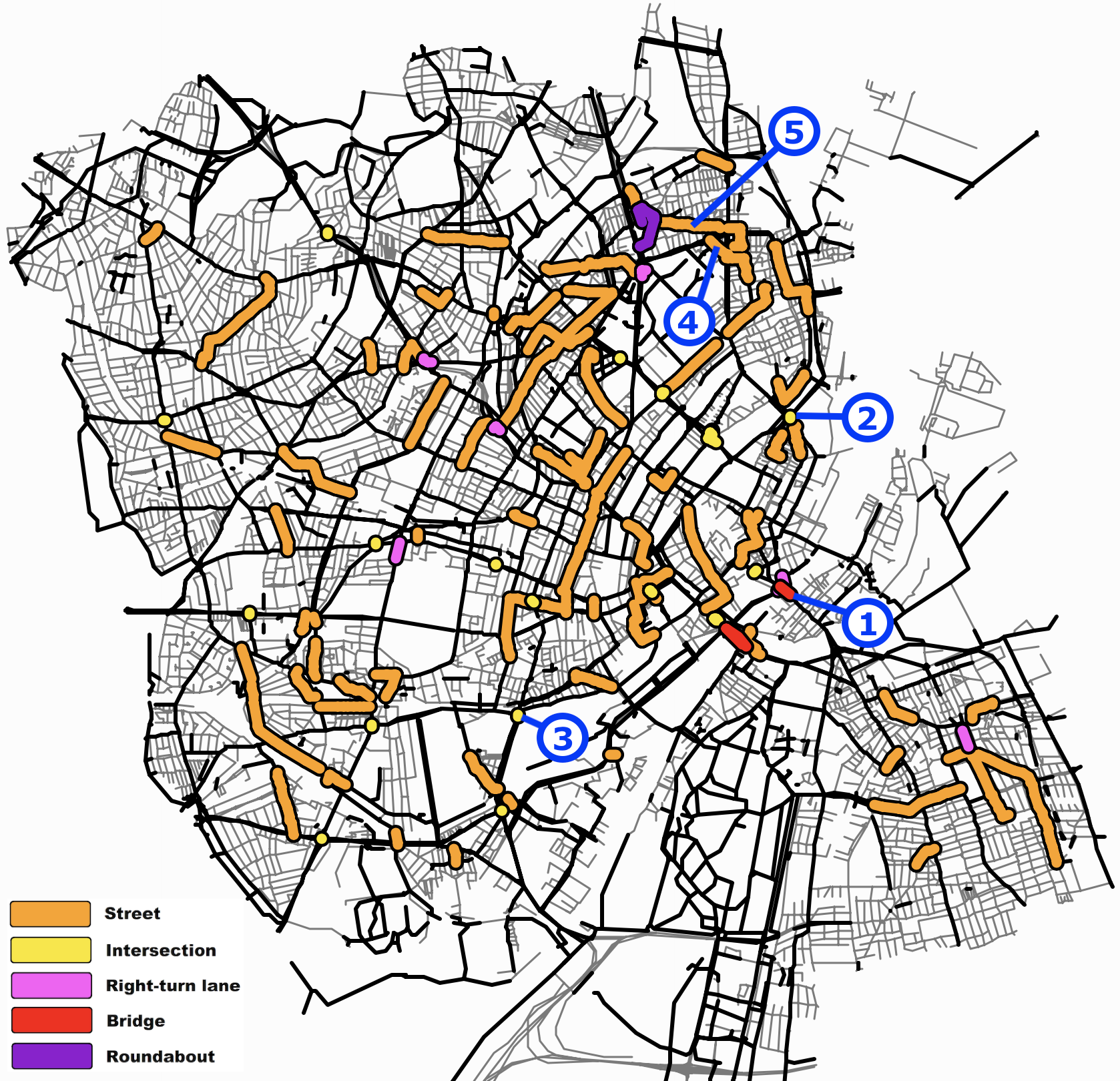} 
    \caption[Gap classes: Overview map of top 105 gaps]{\textbf{Overview map of top 105 gaps by class:} streets in orange, intersections in yellow, bridges in red, right-turn lanes in pink, roundabouts in violet. Errors are shown in \emph{Appendix B}. Numbered blue circles indicate the top 5 gaps (see Fig.~\ref{fig:classes_all_detail} for detail plots). The street network is shown in grey, the bicycle network in black. See \url{https://fixbike.net} for an interactive version.}
    \label{fig:classes_all_overview}
\end{figure}

In this section, we discuss the results of the IPDC procedure applied to the use case of Copenhagen. From the list of 134 gaps that were used as input for the last classification step of the IPDC procedure, we discarded 29 gaps classified as errors. We confirmed and classified the remaining gaps through manual inspection and on-site visits and obtained a list of 105 top priority gaps, which is the final result of the IPDC procedure applied to Copenhagen. The distribution of gap classes in the top 105 gaps is reported in Table~\ref{tab:classes_dist} (class \emph{error} excluded). The map in Fig.~\ref{fig:classes_all_overview} gives an overview of all 105 gaps, with classes plotted by color. In the next sections, we summarize the results per gap class. A list of all 105 confirmed gaps with detail maps and addresses can be found in the SI and at \url{https://fixbike.net/table}. 

\begin{figure}[t]
\captionsetup[subfigure]{labelformat=empty}
\centering
\begin{subfigure}[t]{0.19\linewidth}
    \centering
    \includegraphics[width=\textwidth]{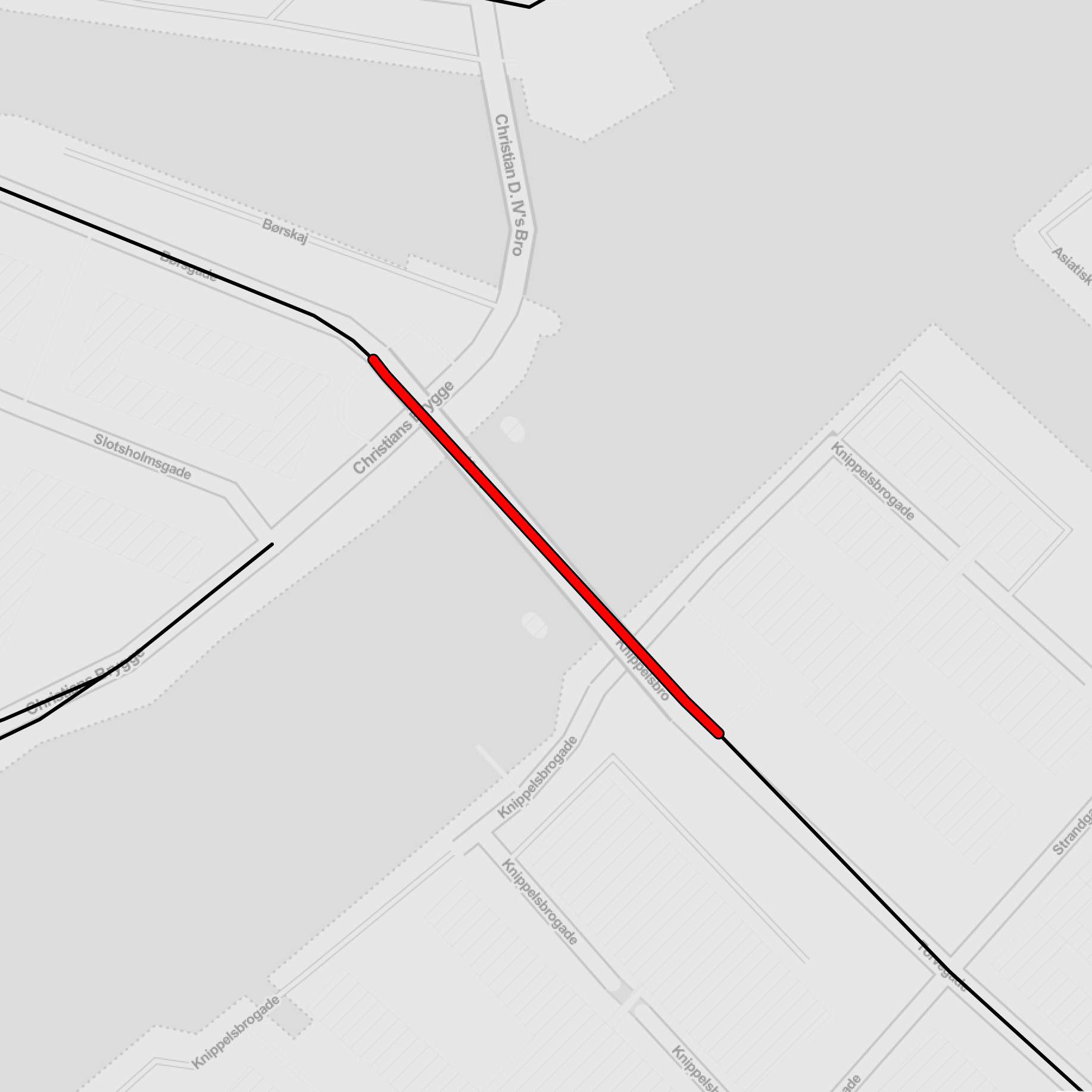}
    \caption{Gap 1}
\end{subfigure}
\begin{subfigure}[t]{0.19\linewidth}
    \centering
    \includegraphics[width=\textwidth]{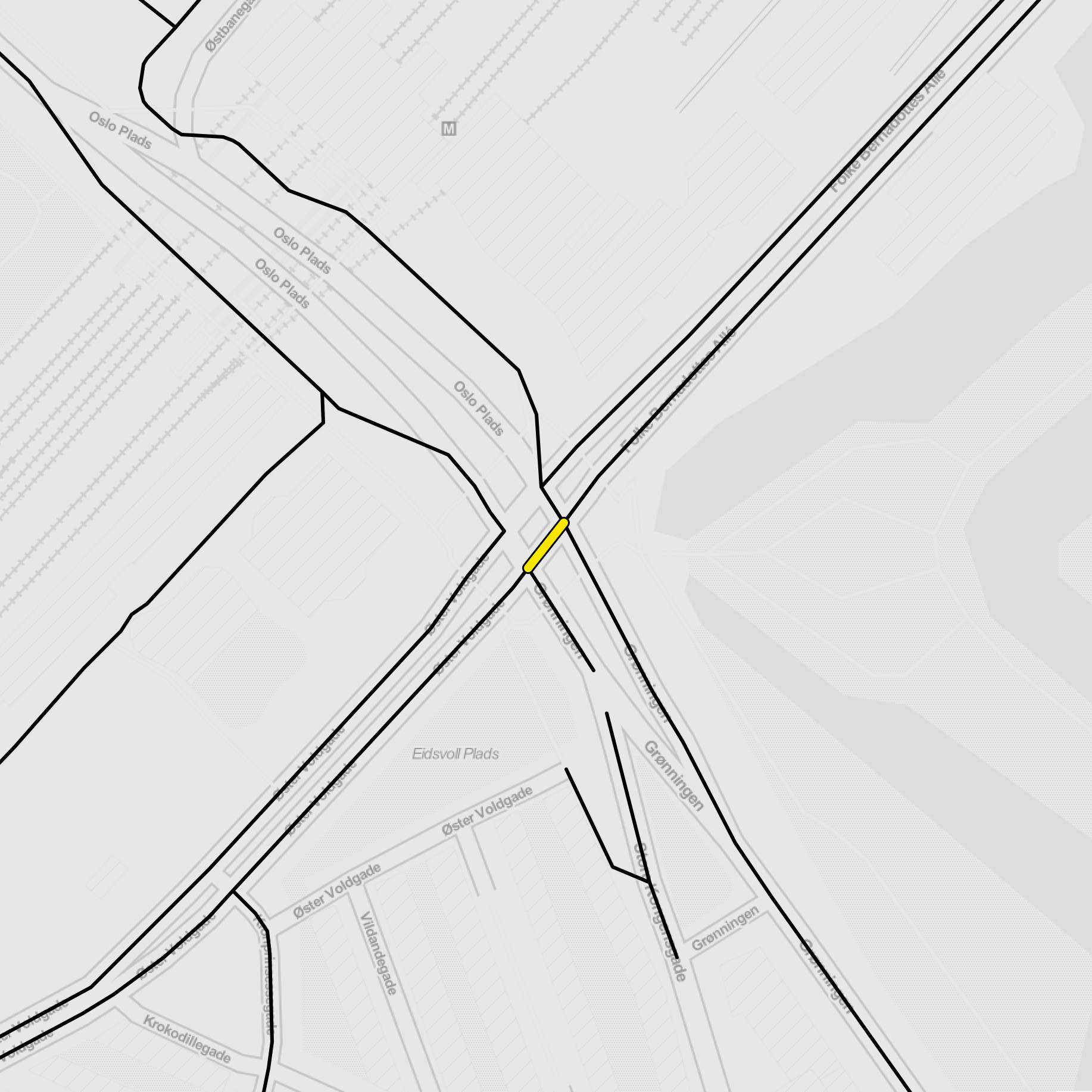}
    \caption{Gap 2}
\end{subfigure}
\begin{subfigure}[t]{0.19\linewidth}
    \centering
    \includegraphics[width=\textwidth]{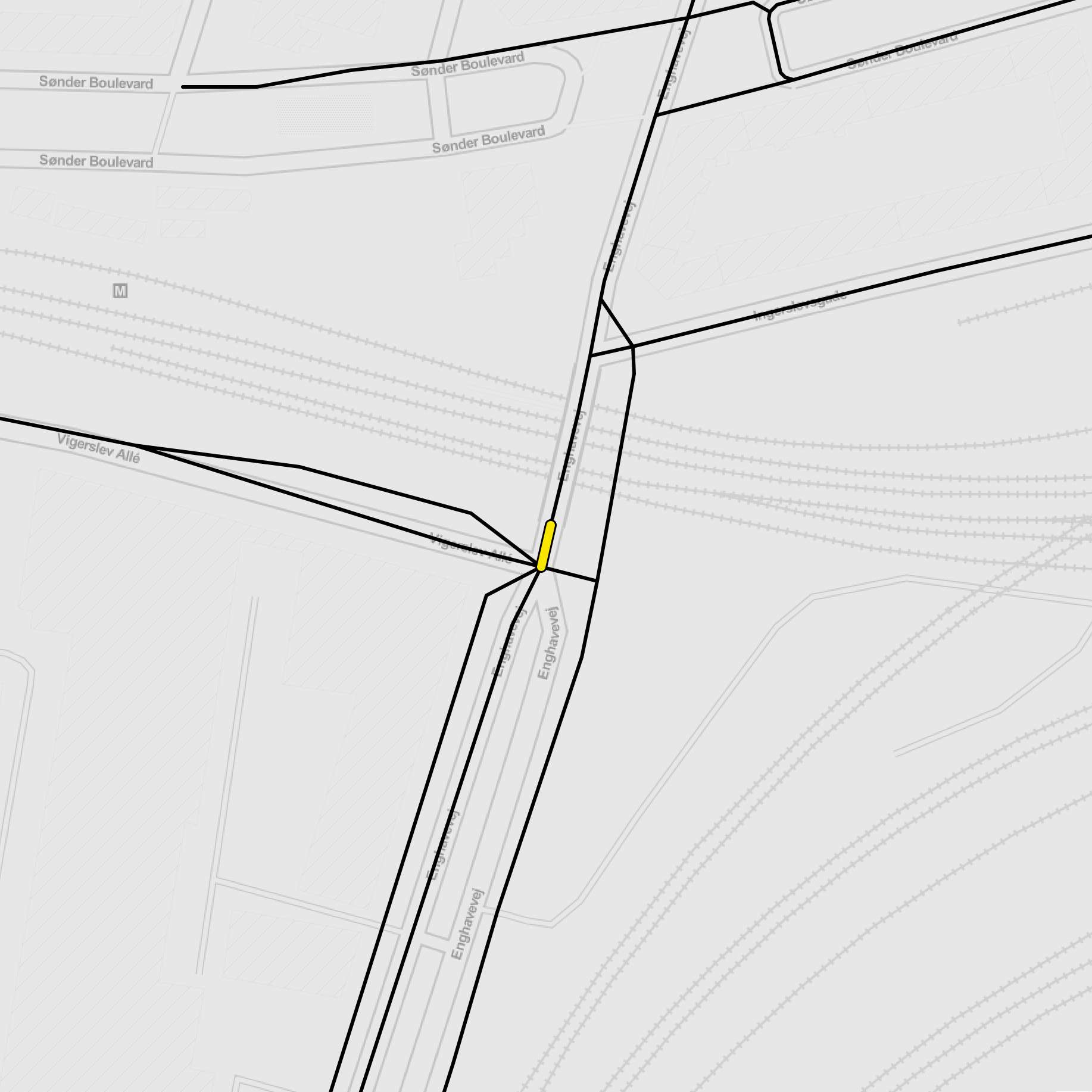}
    \caption{Gap 3}
\end{subfigure}
\begin{subfigure}[t]{0.19\linewidth}
    \centering
    \includegraphics[width=\textwidth]{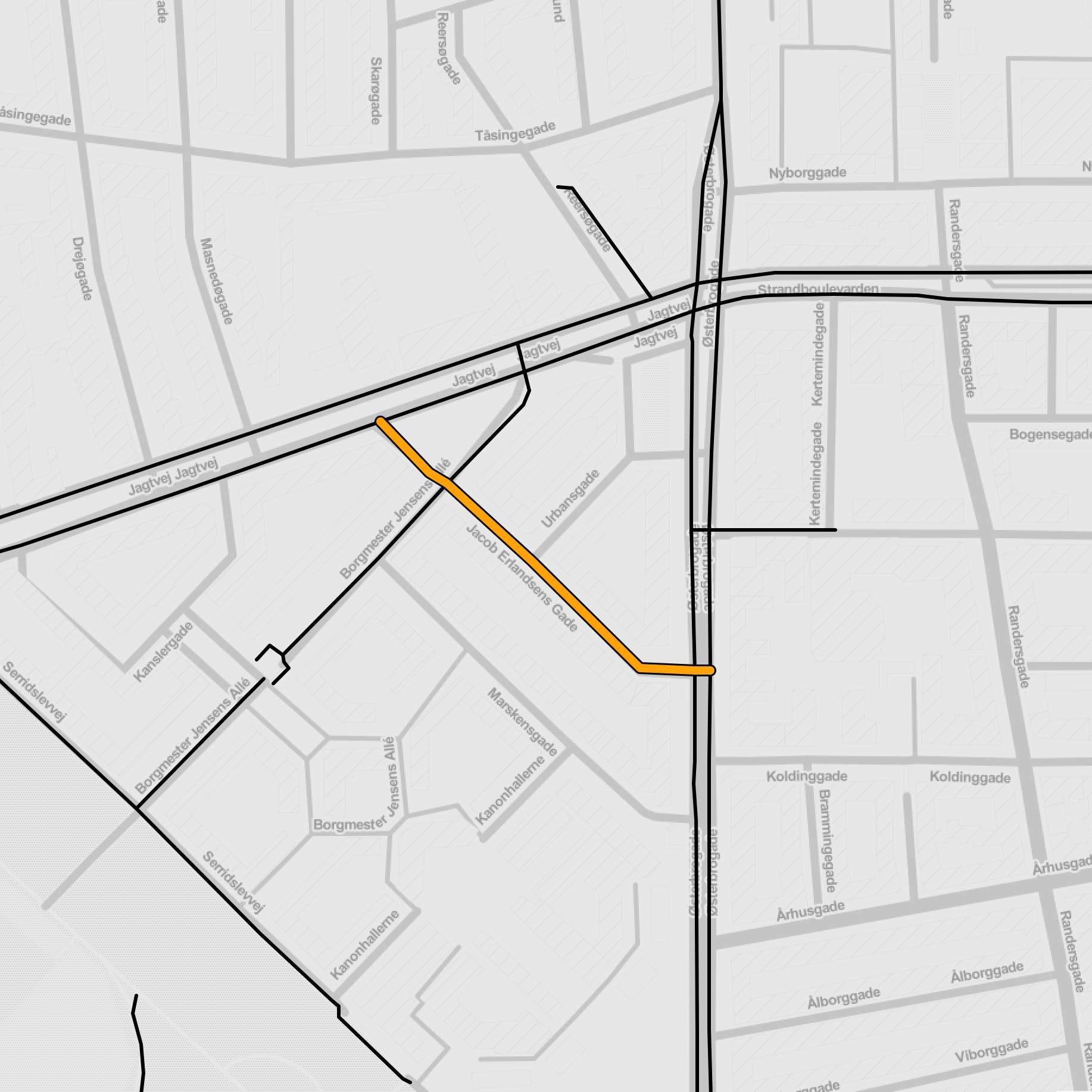}
    \caption{Gap 4}
\end{subfigure}
\begin{subfigure}[t]{0.19\linewidth}
    \centering
    \includegraphics[width=\textwidth]{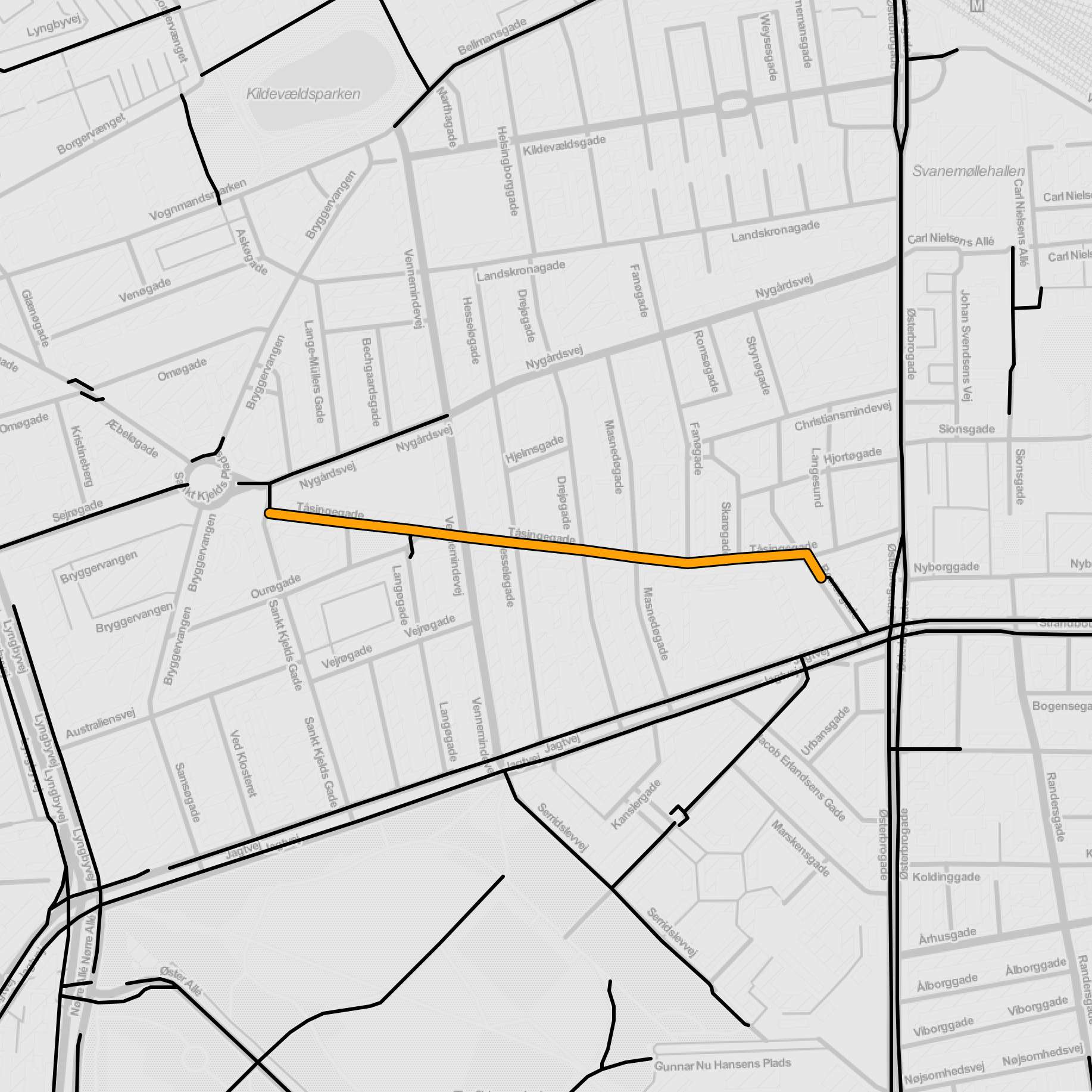}
    \caption{Gap 5}
\end{subfigure}
\caption{\textbf{Detail plots of the top 5 gaps in Copenhagen.} From left to right: Gap 1: Knippelsbro (bridge); Gap 2: Øster Voldgade and Sølvgade (intersection); Gap 3: Enghavevej and Vigerslev Allé (intersection); Gap 4: Jacob Erlandsens gade (street); Gap 5: Tåsingegade (street). All gaps can be explored at \url{https://fixbike.net/table}.}
    \label{fig:classes_all_detail}
\end{figure}

\paragraph{Streets} 
Gaps classified as \emph{street} constitute the majority of our final result (77 out of 105 gaps). Both visual analysis of the gap location and a comparison with Copenhagen's current Cycle Path Prioritization Plan (see section \emph{Comparison with Copenhagen’s Cycle Path Prioritization Plan} below) indicate that several of the identified street gaps might be confirmed as relevant by transport planning practitioners; for example, gap 5 on Tåsingegade (see Fig.~\ref{fig:classes_all_detail}), Gap 17 on Ålandsgade and Frankrigshusene, or gap 23 on Hamletsgade (see SI). Some of the identified street gaps are found on residential streets with presumably low traffic speed and volume, so they would probably not be prioritized from a transport planning perspective in spite of their estimated local relevance indicated by high betweenness values. For example, gap 4 on Jacob Erlandsens Gade (see Fig.~\ref{fig:5gapclasses}) shows that this short street --- although low-traffic --- is an important structural shortcut between the many paths connecting east of Østerbrogade and north of Jagtvej. For a refinement of the procedure, a further distinction of subcategories within the street gap class, both by road conditions (e.g.~speed limit) and by empirical traffic volume data, if available, would be recommended, as it might help to estimate whether these links can be considered safely bikeable in spite of their lack of designated infrastructure \citep{crow2016dmb}.  

Several street gaps come to lie within a locally sparse area of the network and are identified by the IPDC procedure due to the presence of small, isolated bicycle infrastructure elements in their vicinity. Examples are gap 46 on Valløvej and gap 62 on Oxford Allé (see SI). This is a direct consequence of our initial definition of a gap as a path between two bicycle infrastructure elements; hence, in network areas where no bicycle infrastructure at all is present, no gap will be identified, which makes the IPDC procedure less suitable for sparse network areas. 

\paragraph{Intersections} 
With 17 out of the top 105 gaps, intersections are the second most common gap class. As outlined in the section on gap classification, due to both the data structure and data quality issues in OSM, the IPDC list of gaps classified as \emph{intersection} should be understood as a non-exhaustive list of locations where checking for appropriate intersection design is recommended. It is noteworthy that most of the intersections identified by the IPDC procedure also received a considerable number of mentions as ``busy intersections'' in the citizen survey within the Cycle Plan. This is seen, for example, for gap 2 at Øster Voldgade and Grønningen and gap 3 at Enghave Vej and Vigerslev Allé (see Fig.~\ref{fig:classes_all_detail}), as well as gap 65 at H.C.~Andersens Boulevard and Rystensteensgade (see SI).  

\paragraph{Right-turn lanes} 
Seven out of the top 105 maps are classified as \emph{right-turn lane}. Examples are gap 6 at Nørre Allé and Øster Allé (see Fig.~\ref{fig:5gapclasses}), gap 9 at Backersvej and Øresundsvej and gap 14 at Borups Allé and Hillerodgade (see SI). The relatively low number of right-turn lanes in the top 105 gaps identified by the IPDC procedure can partially be explained by tagging inconsistency in OSM, already mentioned above with regard to intersections. We deem it likely that there is a significant number of false negatives, i.e.~right-turns that have not been identified as gaps by the IPDC procedure because they are tagged as ``protected bicycle track'' in OSM. Investigating both the OSM data quality and the objective and subjective safety implications of intersection approach design call for further research. 
 
\paragraph{Bridges} 
There are 3 gaps classified as \emph{bridge} within the top 105 gaps: gap 1 on Knippelsbro (see Fig.~\ref{fig:5gapclasses}) and gaps 8 and 92 on Langebro (see SI). We have already argued for the physical separation of cyclists from motorized vehicles; it is of even higher relevance for the crossing of bridges \citep{melson_influence_2014}. In the case of Copenhagen, bridges play a particularly relevant role as the city is situated on the two islands of Amager and Zealand, and harbours an extensive canal system. According to Copenhagen’s latest Bicycle Account, 7 of the top 10 most heavily trafficked cycling stretches in the city are bridges \citep{city_of_copenhagen_technical_and_environmental_administration_bicycle_2019}. The first three stretches on that list are Dronning Louises Bro, Langebro and Knippelsbro. While Dronning Louises Bro is provided with protected bicycle infrastructure, Langebro and Knippelsbro are not. This aligns well with the results of the IPDC procedure, given that both Langebro and Knippelsbro are listed within the top 105 gaps. The Municipality of Copenhagen is currently in the process of upgrading the cycle lanes on both these bridges to cycle tracks \citep{kobenhavns_kommune_teknik-_og_miljoforvaltningen_cykelsti-prioriteringsplan_2017}. The average benefit of closing a gap classified as \emph{bridge} is almost twice as high as the average benefit for all other gap classes, see table \ref{tab:classes_dist}. This insight is in line with the underlying network topology --- such ``bridge edges'' in infrastructure networks are important connections between otherwise separated or even disconnected parts of the network and therefore have particularly high betweenness centrality values. 

\paragraph{Roundabouts} 
The only gap from the list of top 105 gaps classified as \emph{roundabout} is gap 41 on Australiensvej/Bryggerivangen and Sankt Kjelds Plads. The gap contains two roundabouts: the bigger one, on Sankt Kjelds Plads, and the smaller one on the intersection of Australiensvej and Bryggervangen. The Sankt Kjelds Plads roundabout consists of only one lane where motorized vehicles and bicycles mix (see Fig.~\ref{fig:5gapclasses}). Same as in the case of intersections, future work might consider the set of all roundabouts in the city of Copenhagen and examine their design from a cyclist safety perspective. 

\paragraph{Errors} 
Out of the 29 gaps that have been discarded as errors within the last step of the IPDC procedure, classification, 15 were parallel paths (discussed in detail in section \emph{Discarding parallel paths}) and 14 were data issues in OSM. Overview plots of all errors are found in \emph{Appendix B}. Many of the parallel paths occur at large intersections or along streets with multiple lanes and bicycle infrastructure on both sides, for example, at Lyngbyvej or at the crossing of Frederikssundsvej and Borups Allé. Several parallel paths coincide with some of the busiest bicycle corridors in the city, such as Dybbølsbro and H.C.~Andersens Boulevard, which is an encouraging observation for the use of betweenness centrality as a proxy for bicycle traffic flow. All data issues were due to missing tags for protected bicycle infrastructure in OSM, leading to the IPDC procedure identifying gaps in locations where protected bicycle infrastructure is already in place. While some of the missing OSM tags correspond to relatively recent construction of infrastructure, others contain infrastructure that dates back more than a decade. 

\subsection*{Comparison with Copenhagen’s Cycle Path Prioritization Plan}

\begin{figure}[t]

    \centering
    \includegraphics[width=0.8\textwidth]{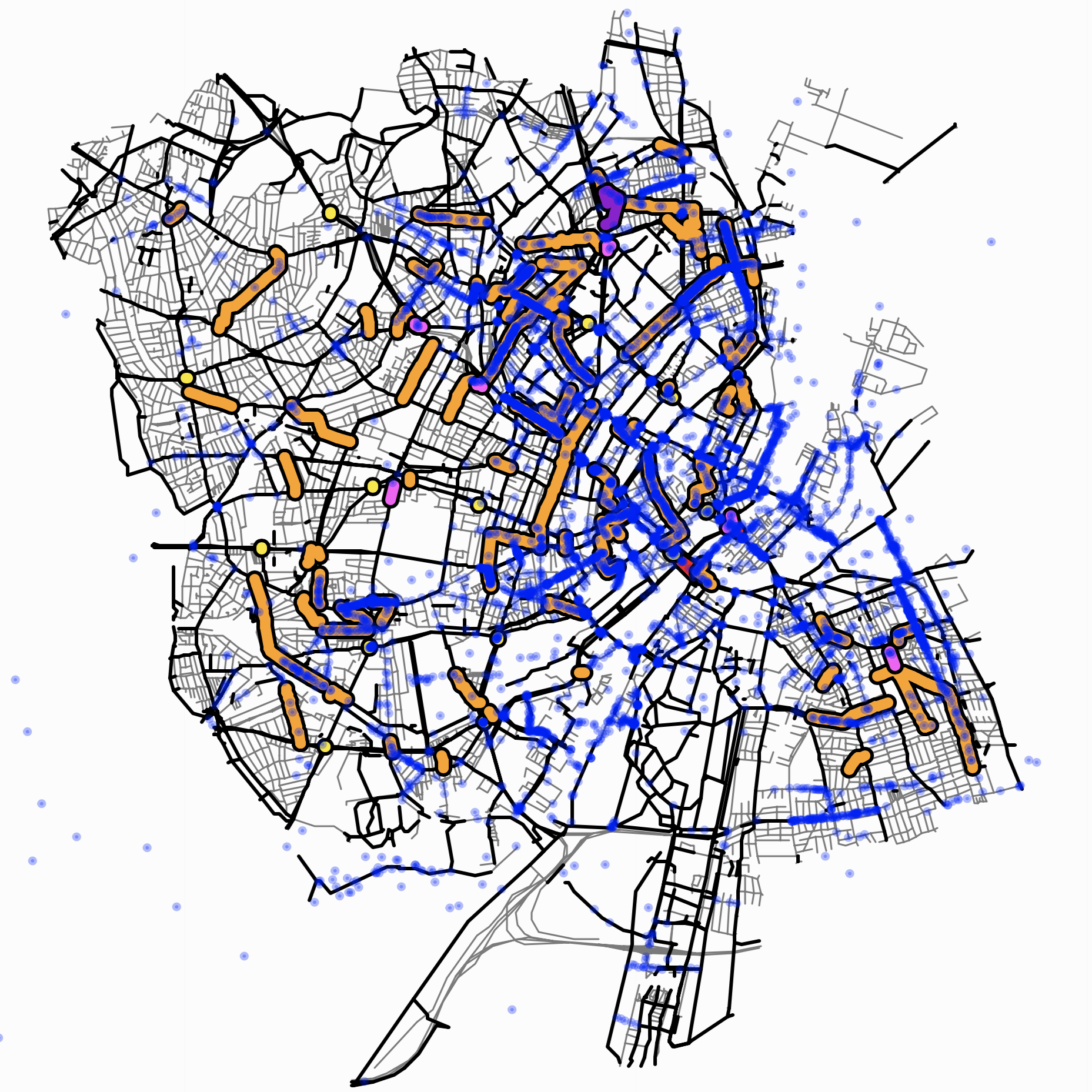}
    \caption{\textbf{Overview map of citizen survey data.} Citizen responses on missing bicycle tracks and busy intersections are represented by blue dots. The street network is shown in grey, the bicycle network in black.}
    \label{fig:ci-overview}
\end{figure}

\begin{figure}[t]
    \centering
    \captionsetup[subfigure]{labelformat=empty}
\begin{subfigure}[t]{0.19\linewidth}
    \centering
    \includegraphics[width=\textwidth]{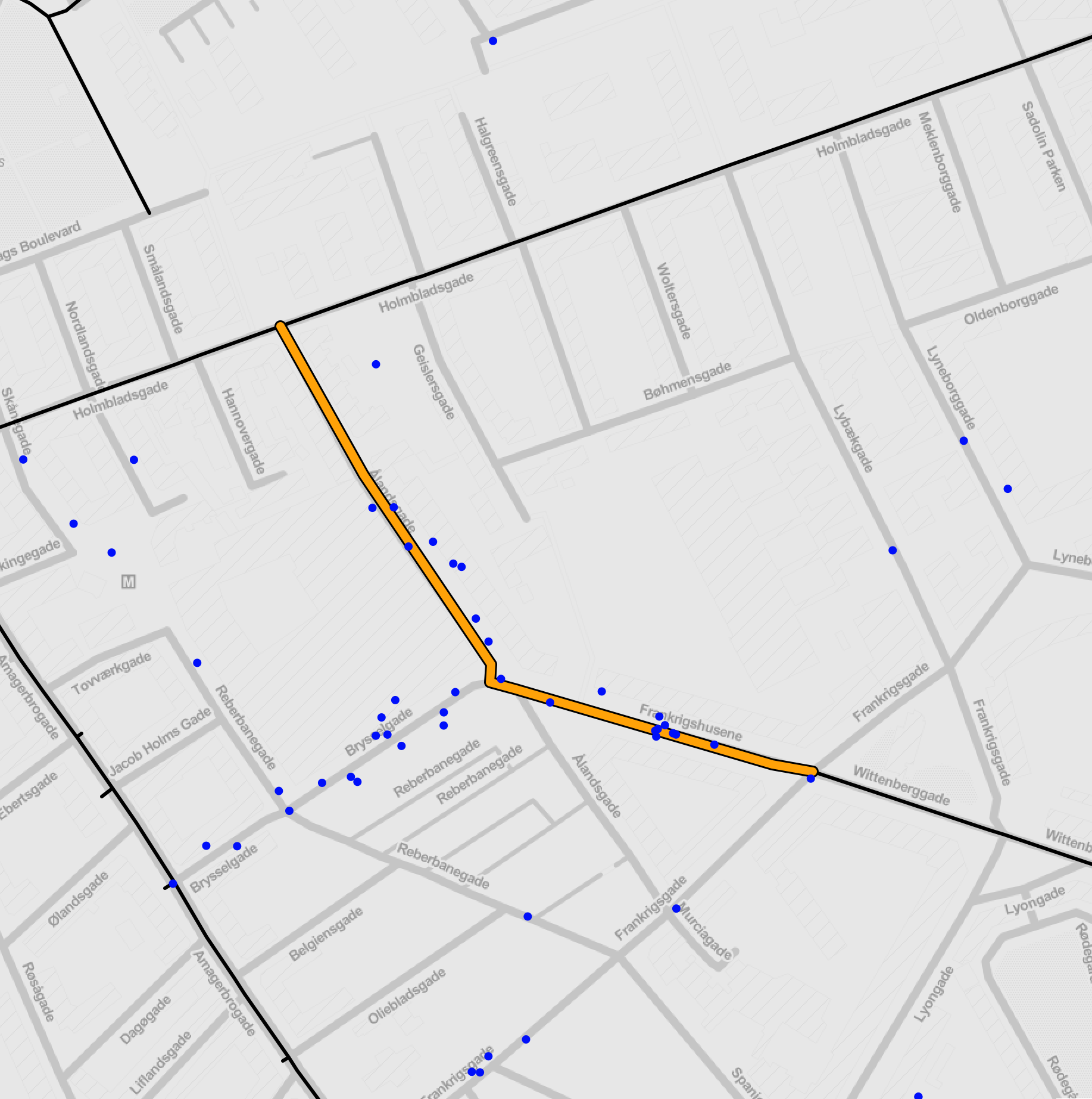}
    \caption{Gap 17}
\end{subfigure}
\begin{subfigure}[t]{0.19\linewidth}
    \centering
    \includegraphics[width=\textwidth]{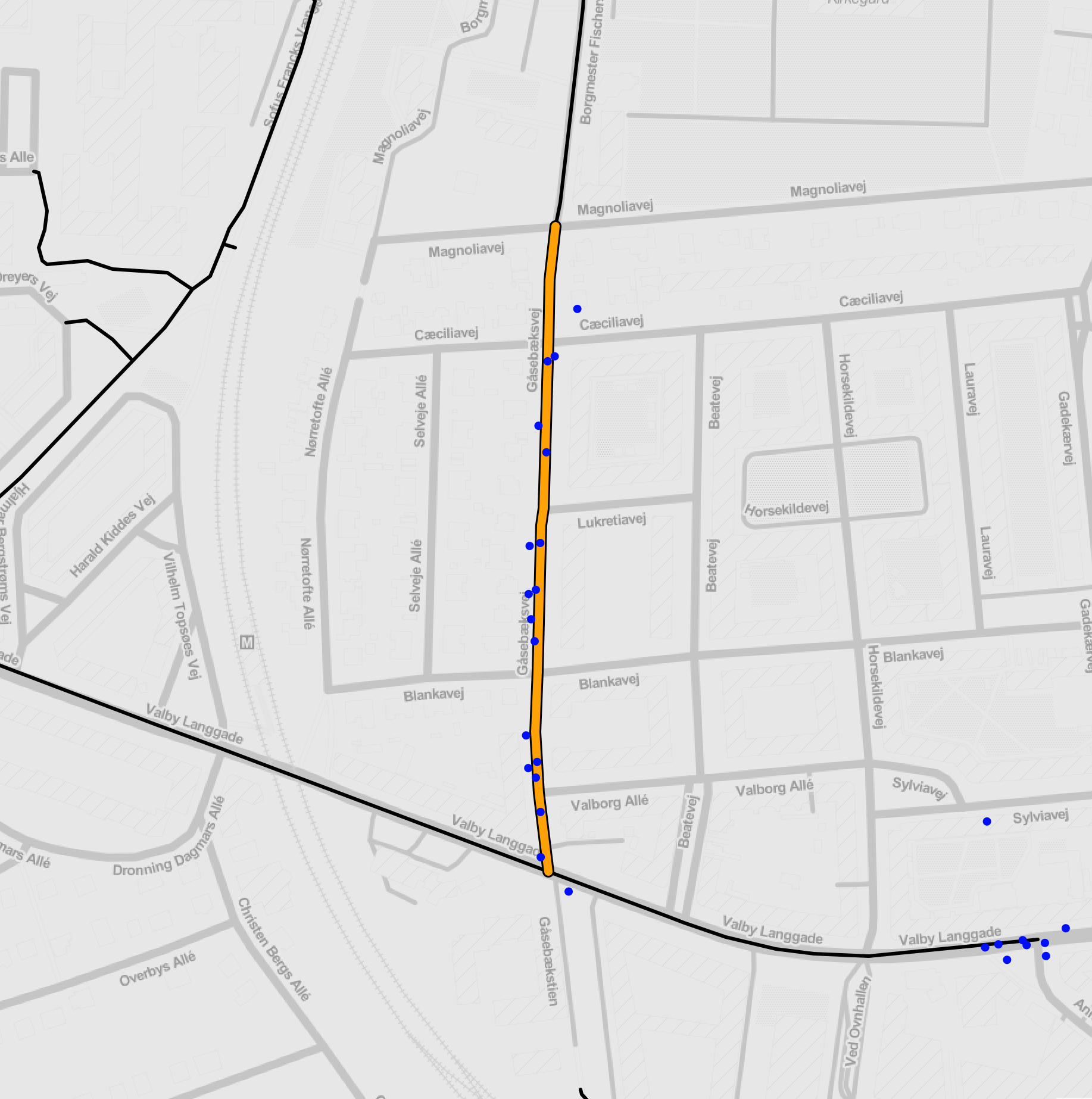}
    \caption{Gap 30}
\end{subfigure}
\begin{subfigure}[t]{0.19\linewidth}
    \centering
    \includegraphics[width=\textwidth]{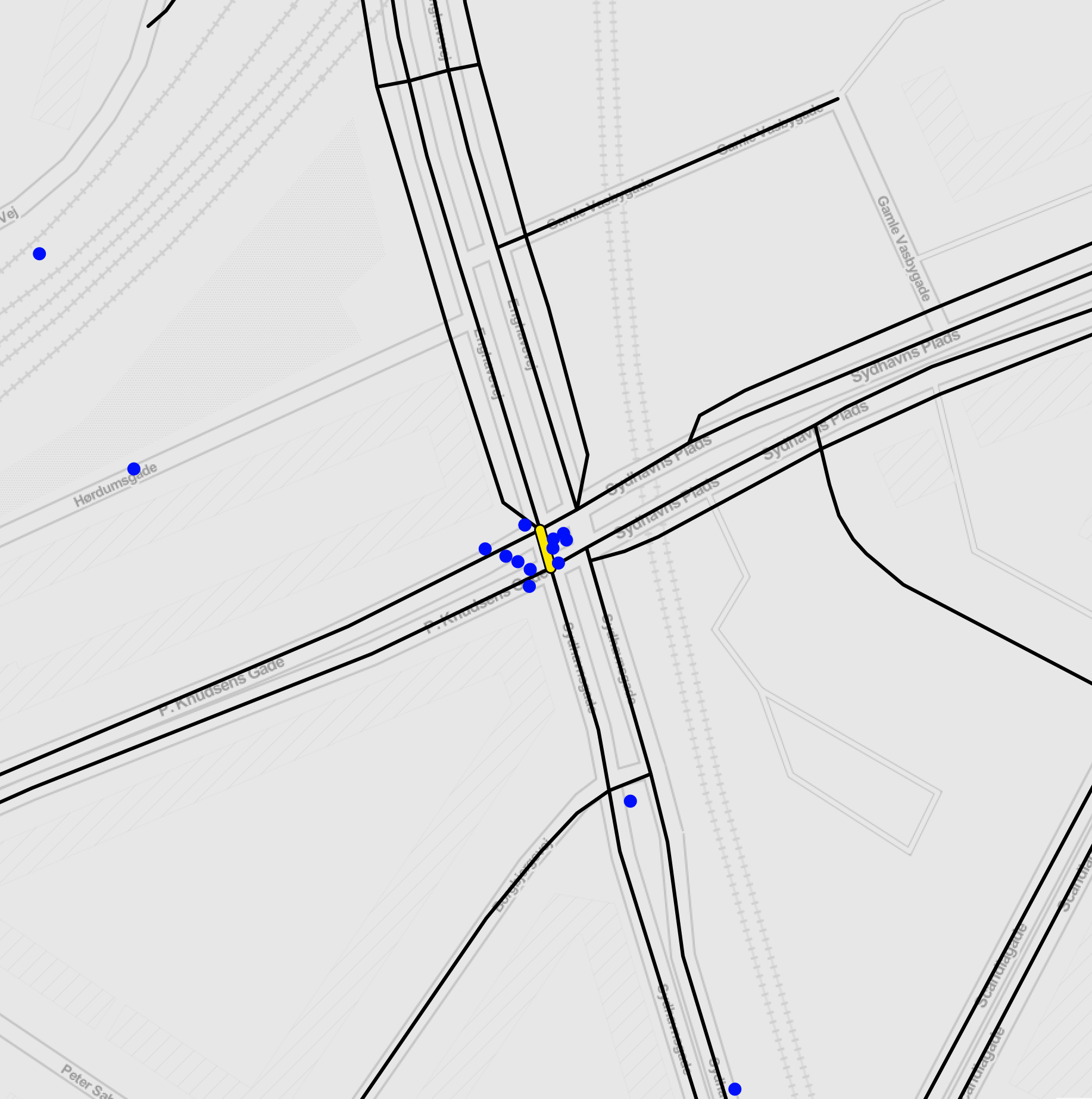}
    \caption{Gap 47}
\end{subfigure}
\begin{subfigure}[t]{0.19\linewidth}
    \centering
    \includegraphics[width=\textwidth]{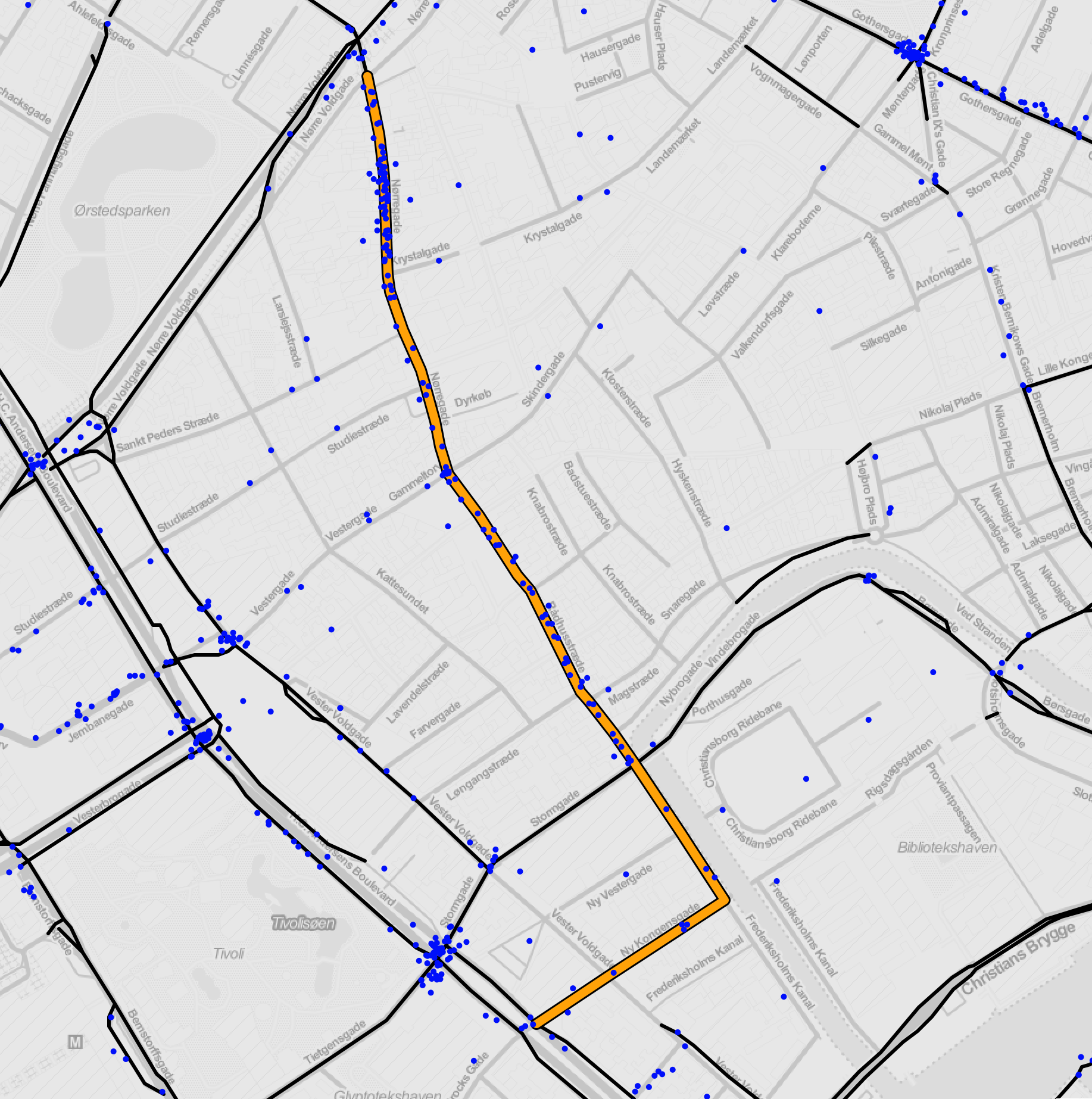}
    \caption{Gap 69}
\end{subfigure}
\begin{subfigure}[t]{0.19\linewidth}
    \centering
    \includegraphics[width=\textwidth]{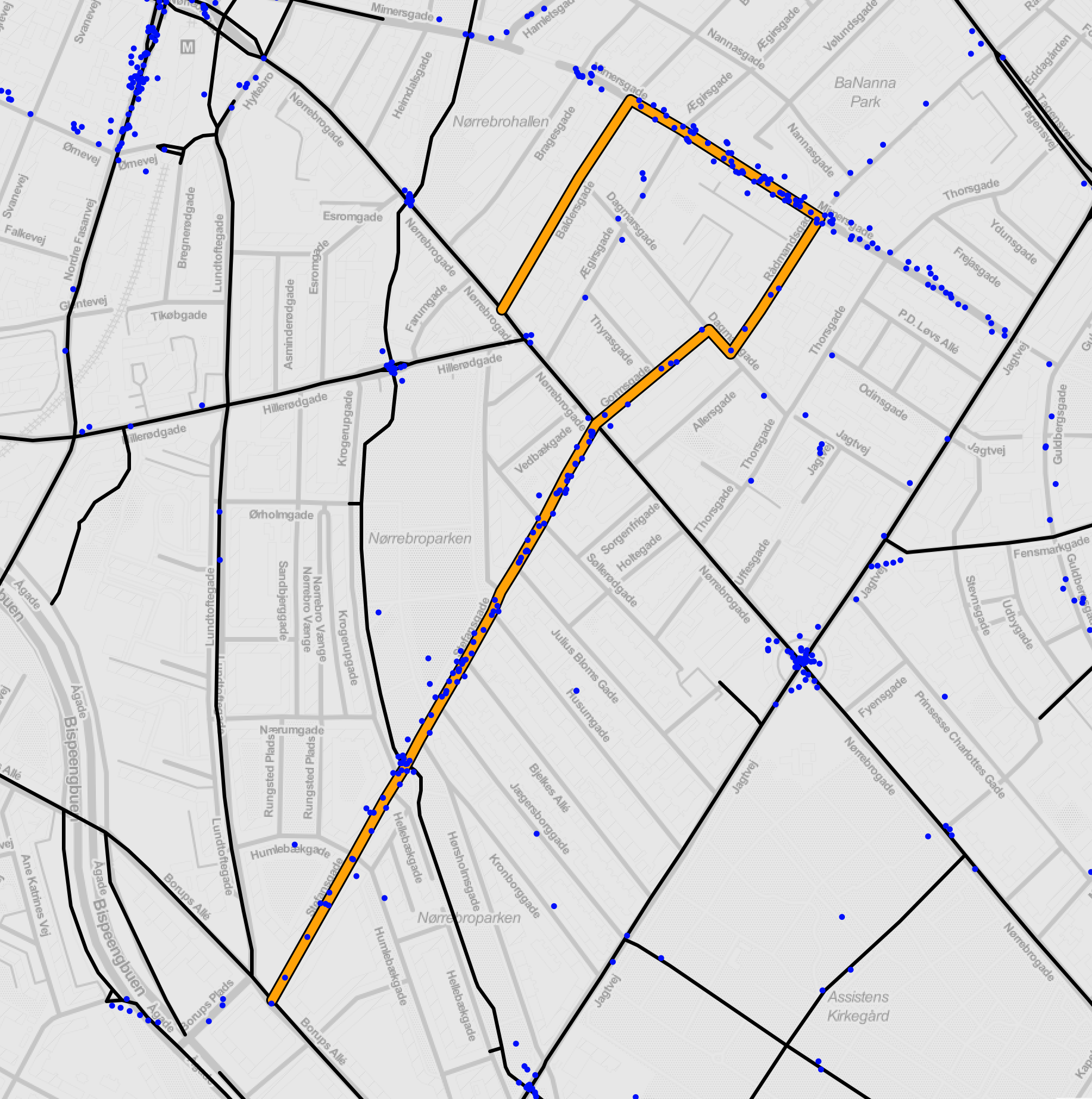}
    \caption{Gap 90}
\end{subfigure}
    \caption{\textbf{Five examples of overlaps between gaps found by the IPDC procedure and citizen survey results (blue dots).} From left to right: Gap 17 on Ålandsgade and Frankrigshusene; Gap 30 on Gåsebæksvej; Gap 47 on the intersection of Enghavevej with P. Knudsens Gade; Gap 69 on Nørregade/Rådhusstræde/Ny Kongensgade; Gap 90 on Stefansgade/Gormsgade/Mimersgade.}
    \label{fig:ci-detail}
\end{figure}

The Municipality of Copenhagen's Technical and Environmental Administration (\textit{Teknik- og Miljøforvaltningen}) regularly publishes a Cycle Path Prioritization Plan (\emph{Cykelstiprioriteringsplan}, hereafter referred to as Cycle Plan). The current plan for the period 2017--2025 \citep{kobenhavns_kommune_teknik-_og_miljoforvaltningen_cykelsti-prioriteringsplan_2017} contains an overview of planned infrastructure improvements and measures targeted at increasing the modal share of cycling, split into five categories: new bicycle infrastructure (tracks, lanes, sharrows), improved intersection design, improvements of the Super Cycle Paths (\textit{Supercykelstier}) network, improvements of the Green Cycle Routes (\textit{Grønne cykelruter}) network, and finally, widening of existing cycle tracks. We conducted a comparative analysis of our list of top 105 prioritized gaps with the planned infrastructure improvements listed in the Cycle Plan across all categories except the last one (given that the width of bicycle infrastructure was not considered in this study). The comparison shows a considerable overlap, given that 46 out of 105 gaps identified by the IPDC procedure are found in locations that are also prioritized in the Cycle Plan. There is a particularly good overlap with the list of high priority routes for new cycle tracks (\textit{højt prioriterede strækninger til nye cykelstier}): 19 out of 35 prioritized routes are included in our list of top 105 prioritized gaps \citep[p. 17]{kobenhavns_kommune_teknik-_og_miljoforvaltningen_cykelsti-prioriteringsplan_2017}. Further categories that show considerable overlap are the list of cycle lanes to be upgraded to cycle tracks (coinciding with 11 of our gaps); as well as identified missing links and planned upgrades of the Green Cycle Routes network (coinciding with 10 of our gaps). 

The Cycle Plan also contains the results from a citizen survey on bicycle infrastructure improvements, conducted by the Municipality of Copenhagen in September and October 2016 \citep{kobenhavns_kommune_teknik-_og_miljoforvaltningen_cykelsti-prioriteringsplan_2017}, which we used for a further qualitative assessment of the present study. Results from the citizen survey consist of a set of geocoded locations, indicated by respondents through clicking on a digital map, for each of the following categories: \textit{Cykelsti mangler} (cycle track missing), \textit{Cykelsti for smal} (cycle track too narrow) and \textit{Kryds med stor traengsel} (busy intersection). We did not utilize responses from the category on too narrow cycle tracks, given that street width was not accounted for in the present study. Figure~\ref{fig:ci-overview} provides an overview of the processed data from the citizen survey. 

A qualitative comparison of our list of top 105 prioritized gaps with the citizen survey results shows considerable overlaps at several locations. Examples are shown in Fig.~\ref{fig:ci-detail}. In total, 71 out of our 105 gaps have at least one mention in the considered categories of the citizen survey. Although these overlaps are encouraging at first glance, there is a relevant caveat to consider. While participatory approaches can improve the equity impact of transportation plans \citep{boisjoly_opening_2017}, a failure to adequately design them might introduce biases and undermine the applicability of the findings \citep{schonlau_selection_2009, nohr_how_2018}. A reliable survey design should account for several bias/equity considerations, such as survey language, medium used, distribution channels and socio-demographic variables of respondents. We have no information about such considerations (or the lack thereof) for the survey data at hand. Therefore, if a location has no mentions in the citizen survey, it cannot be concluded that the infrastructure is already satisfactory there --- it might be due to an undersampling of residents from that area. As long as such considerations in the citizen survey design are unclear, its results should not be regarded as reliable ground truth.

The partial overlap of our top 105 gaps with the locations prioritized by the Municipality of Copenhagen in the Cycle Plan, as well as with the citizen survey results, is a first proof of concept for the IPDC procedure. At the same time, the gaps from our results that do \textit{not} show up in the Cycle Plan are of particular interest for further evaluation, enhancement of methods and decision-making. In a future dialogue with the Municipality of Copenhagen, the results from the IPDC procedure could be scrutinized to find out which gaps are actual missing links in the bicycle network of Copenhagen and possibly will be prioritized in future infrastructure investments; which gaps are less relevant from an urban planning perspective and indicate a necessity to adjust our method (e.g.~by adding information on street type or non-protected bicycle infrastructure to the analysis); and finally, which gaps have been wrongly identified due to data issues in OpenStreetMap. Thus, the comparison with Copenhagen's Cycle Plan demonstrate that the IPDC procedure has the potential to be used as automatized assistance tool and to  successfully complement manual planning processes, while its results can and must be further scrutinized by urban planners.

\section*{Discussion}

Our results from the Copenhagen case study suggest that benefits from bicycle infrastructure improvements for overall network quality are highly variable and location-dependent, with the gap class \emph{bridge} showing the highest average benefits. We also find that network edge effects in transport network analysis might have detrimental implications for the population in the urban periphery, and that future work is needed to mitigate this bias. These findings illustrate the advantages of considering the network as a whole in the analysis, operating on the ``macro level''. In practice, bicycle infrastructure planning often is highly localized and guided by manual decision-making, taking place on a ``micro level''. Our results show that these two approaches should not be seen in competition, but rather as complementary to each other. This is illustrated by the application of the IPDC procedure to bicycle network of Copenhagen and the comparison of the findings with the city’s Cycle Path Prioritization Plan, since potentially relevant results were obtained in spite of minimum data requirements. We are therefore optimistic about the potential of a computational, data-driven macro level approach to decision-making support for bicycle network planning. The IPDC procedure presented in this study is, however, just a first step towards this goal; in the following sections, we discuss the scope and limitations of our approach, as well as further work needed. 

\subsection*{Scope and limitations} \label{sec:applim}
The potentially most substantial limitation for the results of this study is data quality in OpenStreetMap. OSM data is crowdsourced, which allows for the integration of local knowledge and the provision of open source data, but at the same time often leads to data quality issues due to different skill levels within the mapping community and a lack of coherence in tag criteria applications \citep{kaur_automated_2018}. In addition, OSM data quality significantly varies by location \citep{mooney_towards_2010, haklay_how_2010}. A broader quantitative assessment of OSM data quality is still an open research question \citep{yeboah_analysis_2021, jacobs_openstreetmap_2020}, particularly for bicycle infrastructure \citep{ferster_using_2020}. In our results from the IPDC procedure, many of the identified gaps which were discarded as data issues were due to outdated OSM tags, with substantial portions of recently built bicycle infrastructure not yet included in the OSM data. While the number of tagging edits might potentially be used as a workaround for estimating whether the tag is up-to-date \citep{line_investigating_2021}, ideally the implementation of new bicycle infrastructure elements would go hand in hand with the corresponding update in OSM. Another issue is the lack of coherence in bicycle infrastructure tagging. For example, right-turn lanes where the bicycle track merges with a car lane are sometimes marked as protected bicycle infrastructure; the same goes for unprotected intersections which separate two stretches of protected bicycle infrastructure. Therefore, the definitions of bicycle infrastructure categories within OSM \citep{openstreetmap_contributors_bicycle_2021} might be scrutinized from the viewpoint of intelligibility in order to enhance correct and coherent identification of bicycle infrastructure by mappers across differing local contexts \citep{ferster_using_2020}.

A further limitation consists in our simplified conceptualization of street and bicycle networks based on protected bicycle infrastructure availability. Our study considers only protected bicycle infrastructure as part of the bicycle network, but unprotected bicycle infrastructure can also be an adequate design solution under certain conditions \citep{crow2016dmb}. By binarizing street categories into ``protected'' and ``unprotected'', we assume that it is equally undesirable to cycle on any of the car-only streets, whereas in reality the propensity to cycle in mixed traffic highly depends on such factors as road type, traffic flow, and number of lanes. While the IPDC procedure for Copenhagen delivered relevant findings in spite of these simplifications, results could be further scrutinized by enhancing the network model through a more fine-grained differentiation of road and bicycle infrastructure types. The level of detail that can be introduced into the network model will depend on the level of data availability.

Similarly, both the calculation of the benefit and the estimation of bicycle traffic flow within the IPDC procedure could be enhanced in case corresponding data is available. In this study, we assumed minimum data availability and estimated traffic flows and construction costs based only on topological network properties. This assumption was followed intentionally, since our aim was to develop a general method. However, if empirical traffic flow measurements, origin-destination tables, census data etc.~are available \citep{olmos2020dcf}, the calculation of a flow centrality, construction costs and total benefit for each network element could be made more accurate. A related caveat in relation to betweenness centrality is the finite $\lambda$ parameter that we introduced, as a first attempt to partially mitigate the bias towards the network center. The equity implications of centrality metrics have only recently started to be discussed and there is a knowledge gap regarding their quantification \citep{jafino_transport_2020, jafino2021equity, yamaoka_local_2021}. A systematic analysis of such network edge effects would therefore be of high relevance. However, it also goes beyond the scope of the present study, so future work in that regard is urgently called for.

A further simplification is the assumption that cyclists always choose the shortest path from A to B. This is implied in our definition of betweenness centrality, since the shortest path computations on the network are performed with link weight set equal to link length. Several previous studies have accounted for cyclist preferences in shortest paths computations by providing links with a weighting factor that is based on additional features which quantify link attractiveness for cyclists \citep{broach_where_2012, furth_network_2016, cervero_network_2019,  boisjoly2020bnp}. While such an approach may result in more realistic cyclist flow estimations and mitigates the parallel paths problem for some locations on the network, it comes at a considerable cost: A feature-based weighting factor for network links is highly context-dependent, based on potentially subjective cyclist preferences, and constitutes an additional parameter with a non-trivial impact on the shortest path calculations. We therefore explicitly decided not to consider link weighting factors other than link length for our shortest path computations, but extending our model in this respect would be straightforward.

Lastly, given that the classification scheme presented in this study was derived from a qualitative analysis of results for Copenhagen, it might need to be modified for other local contexts. The same goes for the parameters \(D_{\textrm{min}}\) (minimum detour) and $B(\mathbf{g})$ (cut-off benefit) which have been selected for Copenhagen. Appropriate values for both parameters have been derived empirically, but no statement can be made concerning appropriate parameter values for other cities. Although these two thresholds were selected manually, we do not expect this to affect the robustness of our results -- rather, we expect an adjustment of thresholds to mostly impact the number of gaps found.

\subsection*{Future research}
Based on the findings from this study, we anticipate four major lines of future research. First, the IPDC procedure presented here should be further improved. As discussed in the previous section, \emph{Scope and limitations}, there are numerous ways to make the IPDC procedure more accurate, including a testing of its applicability to other locations. Second, we call for the urgent development of a solid computational basis for data-driven bicycle network planning, following recent first steps \citep{olmos2020dcf, nateraorozco2020dso, mahfouz2021rsp}. We deem it particularly relevant to consider multimodality and the multiplex transport network of a city as a whole \citep{nateraorozco2021mum}, and to include equity considerations as an integral part of the network analysis process \citep{gossling2016utj, pereira_distributive_2017, jafino2021equity}. Third, we emphasize the importance of bicycle infrastructure data quality, availability, and coherence \citep{ferster_using_2020}. Access to high quality data is a necessary precondition to provide a scientific basis for any substantial systemic shift towards more active mobility. Fourth and lastly, in line with our call for better cycling data and for data-driven planning approaches, we recommend to account for limited data availability and corresponding mitigation options in any future work on bicycle network planning.

\section*{Conclusion}
In this study, we developed the IPDC procedure for identifying, prioritizing, clustering and classifying gaps in urban bicycle networks. Our method is based only on topological network properties and thus has minimal data requirements. We applied the IPDC procedure to the city of Copenhagen and obtained a list of 105 top priority gaps. A comparison of our results with the city’s most recent Cycle Path Prioritization Plan showed substantial overlaps, both with citizen input on missing bicycle network links and with the city’s list of prioritized locations for the construction of new bicycle infrastructure. The IPDC procedure demonstrates how data-driven network analysis on a city-wide scale can meaningfully complement manual planning processes. We therefore consider this study a further crucial step towards a consolidation of computational methods for bicycle network analysis. 

\section*{Acknowledgements} 
The authors would like to thank Ahmed El-Geneidy, Ane Rahbek Vierø, Kim Sneppen, Marie Kåstrup and Robin Lovelace for their valuable input and comments, and Københavns Kommune for providing us with the Cycle Path Prioritization Plan data. We gratefully acknowledge the open source data and software that this article is based on: Map data copyrighted by OpenStreetMap contributors and available from \url{https://www.openstreetmap.org}; map tiles by Stamen Design, under CC BY 3.0; and images from Mapillary licensed under Creative Commons Share Alike (CC BY-SA) 4.0. 

\section*{Data and code availability}
The code for the IPDC procedure, as well as the OSM data used as input for the Copenhagen case study, is available on GitHub:~\url{https://github.com/anastassiavybornova/bikenwgaps}.

\printendnotes

\bibliography{references}

\clearpage 

\section*{Appendix A: Data acquisition and processing}

We describe the details of data acquisition and processing, as carried out in the present study, within the following subsections: Data source and data structure; Simplification of OSM data; Representation of intersections in OSM data; and lastly, Declustering. All code described in this section is available on GitHub:~\url{https://github.com/anastassiavybornova/bikenwgaps}.

\paragraph{Data source and data structure}

For data acquisition and data processing we used Python and OSMnx \citep{boeing2017osmnx}. The main data source is OpenStreetMap. The input for the case study on Copenhagen consists of GIS vector data of geographic objects which together form the street network of Copenhagen (streets and intersections, bridges, roundabouts, parking lots, paths through green areas etc.) Intersections are represented as points with geographic coordinates, and street segments are represented as sequences of points. In our network derived from the data, intersections of street segments are interpreted as network nodes, and street segments are interpreted as links. 

All input data was downloaded from OSM in February 2021 in csv file format. Data sets were acquired separately for two partially overlapping networks, which, when combined, form the street network of the municipalities of Copenhagen and Frederiksberg: the network of car infrastructure and the network of protected bicycle infrastructure, or, in more simple terms, the car network and the bicycle network. The limits of the two networks coincide with municipality boundaries, which introduces a cut into the continuous fabric of the street network of the Greater Copenhagen area (see the discussion on network edge effects in the section \emph{Gap prioritization}). For each of the two networks, two data sets were generated through OSMnx: one for the nodes and one for the links. Each node from the data set has the attributes geocoordinates and OSM ID; each link has the attributes geocoordinates, OSM ID, length, street name and oneway/twoway indication, as well as several attributes which have not been used within the scope of this study, such as type of highway and speed limit.

The data on car and bicycle nodes was combined into one data set and the parameter ``node type'' was added. Nodes that appeared only in the bicycle data set were assigned the type ``protected'' and nodes that appeared only in the car data set were assigned the type ``unprotected''. Nodes that appeared in both data sets were assigned the type ``contact''. After this, duplicates were removed. The same procedure was applied to the car and bicycle link data sets: they were merged into one data set with the ``link type'' parameter set to ``unprotected'' (if the link appeared only in the car data set) or to ``protected'' (if the link appeared in the bicycle data set or in both data sets). Duplicated links, i.e.~links with same length and type but opposite origin/destination nodes, were removed.

A graph object was created from the resulting data set using the Python's networkx library. The resulting network had 77 disconnected components, out of which only the largest connected component was kept, while all other disconnected components were dismissed as negligible for the sake of simplicity. In the real street network of the city, disconnected components, i.e.~street segments that are not accessible from any other street segment, are quite rare. The appearance of disconnected components in our data set is mostly due to data quality issues, e.g.~missing street segments that should have been classified as protected links. 

\paragraph{Simplification of OSM data}
Within OSM data, prior to further processing, a curved street is represented by a sequence of several points in geocoordinates, which are connected by straight lines. We shall call the corresponding degree-two nodes, which are introduced only for the sake of preserving the physical shape of a link, ``auxiliary''. The presence of auxiliary nodes in the data set strongly biases the degree distribution of the network towards \(d = 2\). The network can be simplified by replacing a sequence of straight links and their corresponding auxiliary nodes by a single polygon link, while preserving the data on length and coordinates of the aggregated links. OSMnx has a built-in function to export already simplified data sets. For our purposes, however, the simplification had to be carried out on the combined network of protected and unprotected links (as opposed to separately simplifying the car and bicycle network, which is an already automated functionality in OSMnx). This is because nodes which are auxiliary in only one of the two networks would otherwise disappear from the data set, and information on connections and partial overlaps between the car and bicycle networks would be lost in case of separate simplification. Therefore, a network was created from the merged data set of protected/unprotected links and protected/unprotected/contact nodes. Then, a simplification algorithm, described in Box \ref{alg:auxiliary}, was applied to the network to remove all auxiliary nodes.

For the data set used in the present study, the simplification algorithm terminates after seven runs; the highest number of auxiliary nodes associated with a link in the final, simplified network is 54. The only degree-two nodes that appear in the data set after simplification are either meeting points of two links of different types or nodes that are kept to represent loops on the network while maintaining the network simple, i.e.~without parallel links. As expected, the degree distribution of the simplified network significantly differs from the original one, shifting from a high to a low percentage of degree-two nodes.

The final outcome of the data preprocessing is the car and bicycle network of Copenhagen, represented by a simple, loop-free, undirected graph with no auxiliary nodes, where each link has two attributes: \textit{type} (``protected'' or ``unprotected'') and \textit{length}, and each node has the attribute \textit{type} (``protected'', ``unprotected'', or ``contact'').

\begin{algorithm}
\SetAlgorithmName{Box}{boxsimp}
\KwIn{\textbf{Input:} Network \(H\) with auxiliary nodes} \\
\KwOut{Network \(H'\) without auxiliary nodes}
\BlankLine
 \While{auxiliary nodes in \(H\)}{
  \For{node in \(H\)}{\If{node degree \(d(n)=2\) {\bf and} links incident on node have the same type}{place node in stack}
 }
 \While{stack is not empty}{
  take random node \(n\) from stack\;
  \eIf{neighbours of \(n\) are neighbours themselves}{
    remove node \(n\) from stack\;
   }{
remove two links incident on node \(n\) from link set of network \(H\)\; 
add new link connecting two neighbours of \(n\) to the link set of network \(H\)\;
set length attribute of new link to sum of lengths of removed links\;
add geocoordinates of removed links to geocoordinate attribute of new link\; 
remove node \(n\) from node set of network \(H\)\;
remove node \(n\) and, if applicable, its two neighbours from stack\;
  }
 }
 }
\caption{Algorithm for removal of auxiliary nodes from the OSM data set}
\label{alg:auxiliary}
\end{algorithm}

\paragraph{Representation of intersections in OSM data}

Within the OSM data structure, intersections of smaller spatial extent appear as single nodes (a node representing the crossing of two streets), while larger ones appear as a set of nodes and links (each node representing the intersect of two or more lanes --- see the example in Fig.~\ref{fig:fx_cluster} where nodes A, B and C are all part of the same intersection). As a rule, but not exclusively, this is the case when at least one of the intersecting streets is bidirectional. Keeping this representation of larger intersections within the data structure allows for the identification of unprotected crossings, that lie on an otherwise protected bicycle track, as gaps in the bicycle network. However, this method of identifying unprotected crossings is by far not exhaustive. This has several reasons. First, due to the data structure, the IPDC procedure does not recognize unprotected intersections that are represented by single nodes in the network model as gaps. Second, even with a clearly outlined set of intersection design criteria at hand which would enable us to discard protected intersections from the gap list, the incoherence of intersection tagging in OSM results in numerous false negatives and false positives: intersections with a protected crossing for cyclists are often tagged as unprotected bicycle infrastructure; intersections without any bicycle infrastructure are often tagged as part of the cycle track they are actually interrupting.

\paragraph{Declustering}
To decluster the partially overlapping gaps identified by the IPDC procedure, we developed a simple declustering heuristic, which is described in Box \ref{alg:declustering}. Within the declustering process, gaps with a benefit metric of at least $B(\mathbf{g})_{\mathrm{min}}$ are combined into a network $C$, after which each disconnected component of $C$ is declustered separately. Declustered gaps are added to the declustered gap list $d$. Gaps that obtain a benefit metric below $B(\mathbf{g})_{\mathrm{min}}$ are discarded. The list of remaining gaps is the output of the declustering heuristic and the input for the next step in the IPDC procedure (classification). 

In the present study, the benefit metric of $B(\mathbf{g})_{\mathrm{min}}=15\,000$ is used as cut-off value, which results in a list of 1199 gaps as input for the declustering heuristic. The gap network $C$ consists of 101 disconnected components (gap clusters). The resulting gap set contains 168 declustered gaps, out of which 34 are discarded due to their lower values of $B(\mathbf{g})<B(\mathbf{g})_{\mathrm{min}}$; the final output $d$ is a list of 134 gaps.

\begin{algorithm}
\SetAlgorithmName{Box}{boxheur}
 \KwIn{\textbf{Input:} List \(c\) of partially overlapping gaps; cut-off benefit metric $B(\mathbf{g})_{\mathrm{min}}$} \\
 \KwOut{List \(d\) of non-overlapping gaps}
 \BlankLine
Remove gaps with $B(\mathbf{g})<(\mathbf{g})_{\mathrm{min}}$ from \(c\) \;
Combine gaps \(c\) into network \(C\) \;
Decompose network \(C\) into a list of disconnected components \(dc\) \;	\For{\(comp\) in \(dc\)}{
\While{\(comp\) is not empty}{
compute all shortest paths between nodes  $n \in comp  | d(n)  \neq 2$ \;
compute benefit metric $B(\mathbf{g})$ for each path \;
find path $p_{\mathrm{max}}$ with highest value $B(\mathbf{g})_{\mathrm{max}}$ \;
add $p_{\mathrm{max}}$ to final gap list \(d\) \;
remove $p_{\mathrm{max}}$ from \(comp\) \;
}

    }
Remove gaps with $B(\mathbf{g})<(\mathbf{g})_{\mathrm{min}}$ from \(d\) 
\caption{Declustering heuristic for overlapping gaps}
\label{alg:declustering}
\end{algorithm}

\clearpage

\section*{Appendix B: Error plot}

\begin{figure}[h]
    \centering
    \includegraphics[width=0.95\textwidth]{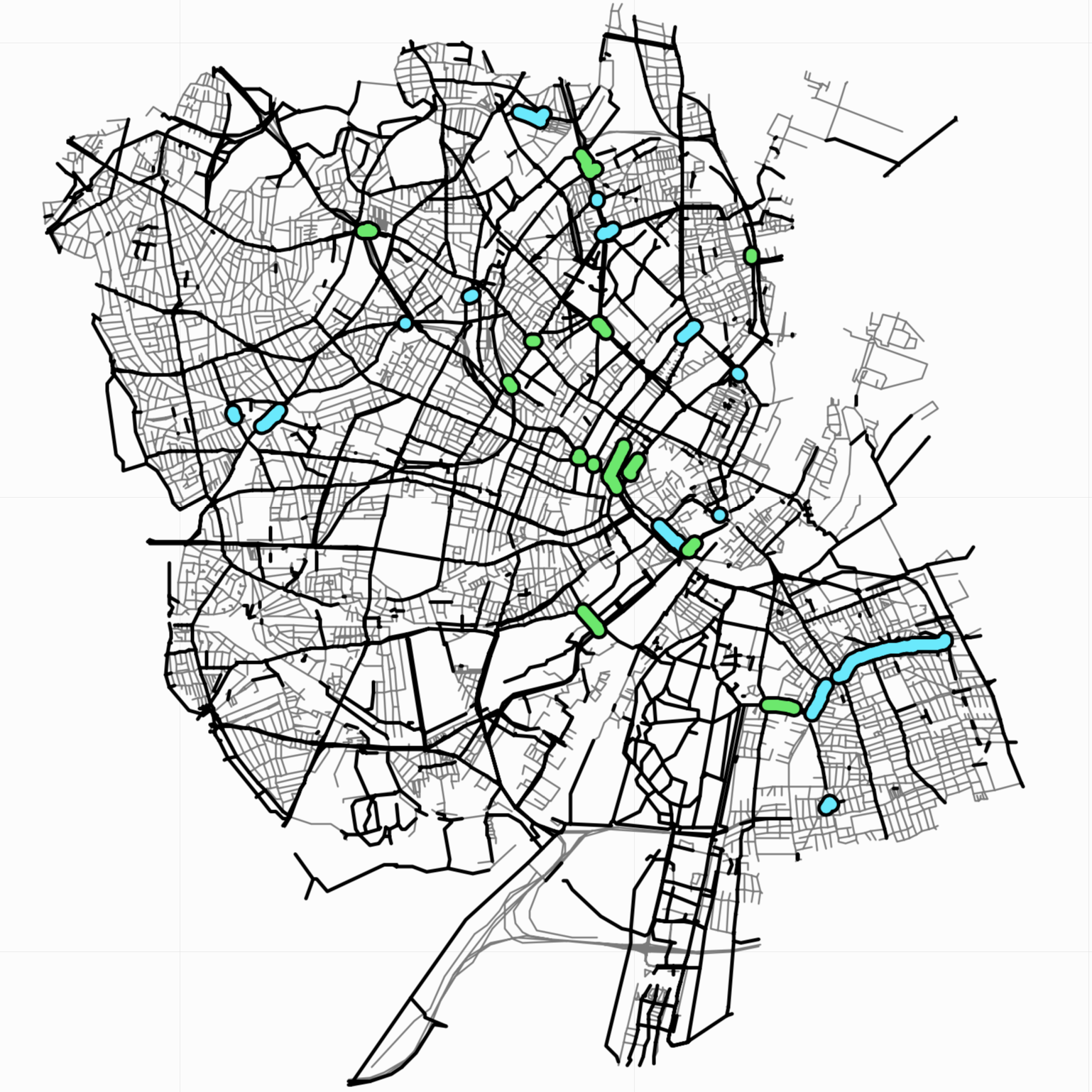} 
    \caption{\textbf{Errors in the list of top ranked gaps in Copenhagen.} 29 out of 134 gaps identified by the IPDC procedure in the Copenhagen network that have been discarded as errors: data issues in light blue; parallel paths in light green.}
    \label{fig:errors_overview}
\end{figure}

\end{document}